\documentclass[conference]{IEEEtran}
\IEEEoverridecommandlockouts

\usepackage{graphicx}

\usepackage{url} 
\usepackage{dashbox} 

\usepackage{amsmath}

\usepackage{pifont} 

\usepackage{lipsum}
\usepackage{float}
\usepackage{fancyref}
\usepackage{color,soul} 
\usepackage{balance}

\usepackage{tabularx}
\usepackage{booktabs}

\usepackage[]{xcolor}
\usepackage{xspace}
\usepackage{pdflscape} 

\usepackage{subcaption}
\usepackage{multirow}
\usepackage{listings}
\usepackage[inline]{enumitem}
\setlist{nolistsep,leftmargin=*}

\newcommand{\tegcer}{TEGCER\xspace}
\newcommand{\tegcerfull}{\textbf{T}argeted \textbf{E}xample \textbf{G}eneration for \textbf{C}ompilation \textbf{ER}rors\xspace}

\newcommand{\tracer}{Tracer\xspace}

\newcommand{\acci}[1]{$\mathsf{Pred}@\mathsf{#1}$\xspace}

\newcommand{\kac}{\ensuremath{\mathsf{k}}\xspace}

\newcommand{\cmark}{\ding{51}}%
\newcommand{\xmark}{\ding{55}}%

\newcommand{\rfill}[1]{\colorbox{red!30}{\!#1\!}}
\newcommand{\gfill}[1]{\colorbox{green!30}{\!#1\!}}

\newcommand{\code}[1]{\texttt{#1}}

\newcommand{\dottedsquare}[1][H]{\setlength{\fboxsep}{0pt}\setlength{\dashlength}{2.2pt}\setlength{\dashdash}{1.1pt} \dbox{\phantom{#1}}}

\newcommand{\etal}{\textit{et al.}}

\begin{document}

\title{Targeted Example Generation for \\ Compilation Errors}

\makeatletter
\newcommand{\linebreakand}{%
  \end{@IEEEauthorhalign}
  \hfill\mbox{}\par
  \mbox{}\hfill\begin{@IEEEauthorhalign}
}
\makeatother

\author{
\IEEEauthorblockN{Umair Z. Ahmed\IEEEauthorrefmark{1}}
\IEEEauthorblockA{
{IIT Kanpur} \\
umair@cse.iitk.ac.in}
\and
\IEEEauthorblockN{Renuka Sindhgatta\IEEEauthorrefmark{1}}
\IEEEauthorblockA{
{Queensland University of Technology} \\
renuka.sr@qut.edu.au}
\and
\IEEEauthorblockN{Nisheeth Srivastava}
\IEEEauthorblockA{
{IIT Kanpur} \\
nsrivast@cse.iitk.ac.in}
\and
\IEEEauthorblockN{Amey Karkare}
\IEEEauthorblockA{
{IIT Kanpur} \\
karkare@cse.iitk.ac.in}

\thanks{\IEEEauthorrefmark{1}Part of this work was carried out by the author at IBM Research.}
}

\maketitle  
\IEEEpubidadjcol

\begin{abstract}
We present TEGCER, an automated feedback tool for novice programmers. TEGCER uses supervised classification to match compilation errors in new code submissions with relevant pre-existing errors, submitted by other students before. The dense neural network used to perform this classification task is trained on 15000+ error-repair code examples. The proposed model yields a test set classification Pred@3 accuracy of 97.7\% across 212 error category labels. Using this model as its base, TEGCER presents students with the closest relevant examples of  solutions for their specific error on demand. A large scale ($N>230$) usability study shows that students who use TEGCER are able to resolve errors more than 25\% faster on average than students being assisted by human tutors.

\end{abstract}

    
    
%


\section{Introduction}

\begin{figure*}[tbp] 
    \centering
        
    \includegraphics[width=.99\linewidth, keepaspectratio]{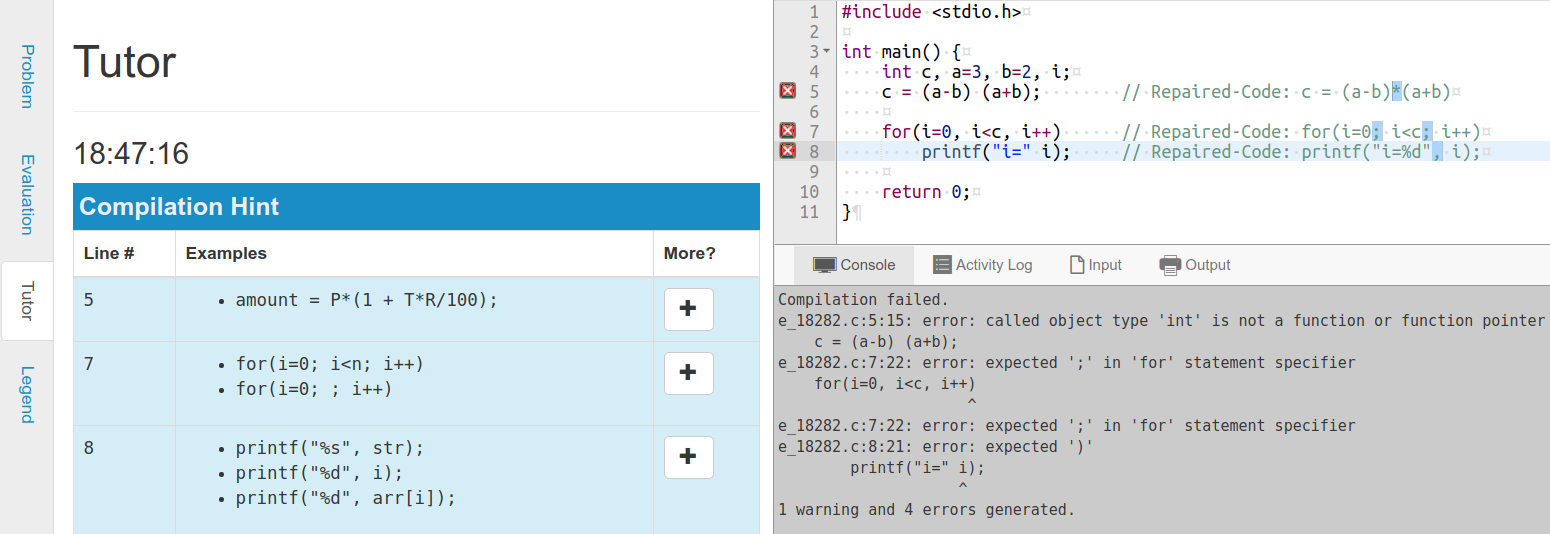}
    \caption{User interface of TEGCER feedback tool. The erroneous student code is written in top-right \textit{editor} pane, and the compilation errors are displayed in bottom-right \textit{console} pane. \tegcer's example feedback is shown in the  left \textit{tutor} pane.}
    \label{fig:eg_ui}
\end{figure*}

CS1, the Introduction to Programming course, is one of the most popular college courses with class sizes reaching $1000+$ students in some universities~\cite{ucbClass}, and up to hundreds of thousands on Massive Open Online Courses (MOOCs)~\cite{udacityClass}. 
Programming assignments are used as an important learning tool to help acquire and practice programming skills. Presently, CS instruction is struggling to cope with the challenge of effectively imparting such programming lessons at scale~\cite{camp2015booming}. In this paper, we present a potential solution to a part of this challenge - providing personalized feedback to correct programming errors.

While attempting programming assignments, students run into compilation errors and logical errors. Compilation errors typically occur when a program does not obey the syntax or grammar of a language, and are flagged by a compiler. However, the associated errors messages are often cryptic and unclear to novice programmers~\cite{mccauley2008debugging,becker2019unexpected}, especially those undertaking their first programming course. In order to help students, enhancements to compiler messages have been proposed~\cite{becker2018effects,becker2019unexpected}, which involves manually analyzing the top-frequent error messages by experts and listing possible solutions, raising scalability concerns.

A large number of repair tools
have been proposed to automatically detect and repair errors in programs, enumerated in the survey paper by Monperrus~\cite{monperrus2015automatic}. The current state-of-art compilation error repair tools are able to achieve up to $44\%$ repair accuracy~\cite{ahmed2018compilation}. These repair tools typically take as input the incorrect solution by a student, also referred to as erroneous or buggy code, and generate the correct repaired code as their final output. While revealing these correct solution could be invaluable as a feedback in certain situations to help student proceed further, it is unclear if providing it during the programming exercise could aid in effective learning. It is plausible for the student to simply copy the answer without understanding the error causation or its fix.

Novice programmers consider example programs as the most useful type of material for learning \cite{lahtinen2005}. The notion of learning by example has been widely studied in education research \cite{sweller1999}. While there is no precise definition, an example aims to convey how another similar problem can be solved. 
In this paper, we propose a new feedback tool called \tegcer (\tegcerfull). The goal of \tegcer is to provide an alternative feedback to students who encounter compilation errors in their program, while maintaining the scalability offered by automated repair tools. This feedback is in the form of examples of fixes performed by other students, albeit on a different program, when faced with similar error previously. Such an approach is a departure from most of the recent work in literature, which instead tend to focus on producing the desired repair/solution~\cite{gupta2018deep,ahmed2018compilation,gupta2017deepfix,bhatia2016automated}, help improve the descriptive error message~\cite{becker2018effects}, or vary the error message's structure and placement~\cite{nienaltowski2008compiler}.

Figure~\ref{fig:eg_ui} presents an erroneous code attempt by student, on our custom IDE Prutor~\cite{das2016prutor} with \tegcer feedback tool integrated. The students code, written in the top-right \textit{editor} pane, suffers from compilation errors in 3 separate lines. In line $\#5$ and $\#8$, the student forgot to use an asterisk operator `\code{*}' and comma separator `\code{,}', respectively. While in line $\#7$, the student mistakenly uses comma separator, instead of the semi-colon separator `\code{;}' required in for-statement specifier. The error messages returned by Clang~\cite{lattner2004llvm}, a popular compiler for C language, are then shown in the bottom-right \textit{console} pane. The error messages for line $\#5$ and line $\#8$ are cryptic, with the compiler treating them as an incorrect function invocation and misplaced bracket, respectively. This can be even more confusing for novice students, learning their first programming language, usually requiring human tutor assistance to understand the error and then fix it.

\tegcer's example feedback is shown in the left-side \textit{Tutor} pane for each individual erroneous line. These automatically generated examples are highly relevant to both, the mistake made by student and its desired repair. The students can request for multiple examples of repaired code, from which they can be expected to learn the general cause of a particular compilation error and its potential fixes, after which they can proceed to transfer the acquired knowledge on their own code and repair it.


There have been extensive studies in literature on how providing relevant worked examples can help students learn effectively in a pedagogical setting~\cite{renkl2014toward}. This idea of using code examples from other students as compilation error feedback is not new. The closest related work to ours is HelpMeOut~\cite{hartmann2010helpmeout}, where students can query a central repository to fetch example erroneous-repaired code pairs of other students, that suffer from compilation errors similar to their own code. This repository of errors is created and maintained manually by students, and relies on user provided explanation and ratings for suggesting relevant examples. 
 In contrast, \tegcer passively learns from mistakes made by students of previous offerings, and provides feedback for students in new future offerings of course.





\begin{figure*}[tb] 
    \centering
    \mbox{ 
        \centering
        \begin{subfigure}[t]{.62\linewidth} 
            \includegraphics[width=\linewidth, keepaspectratio]{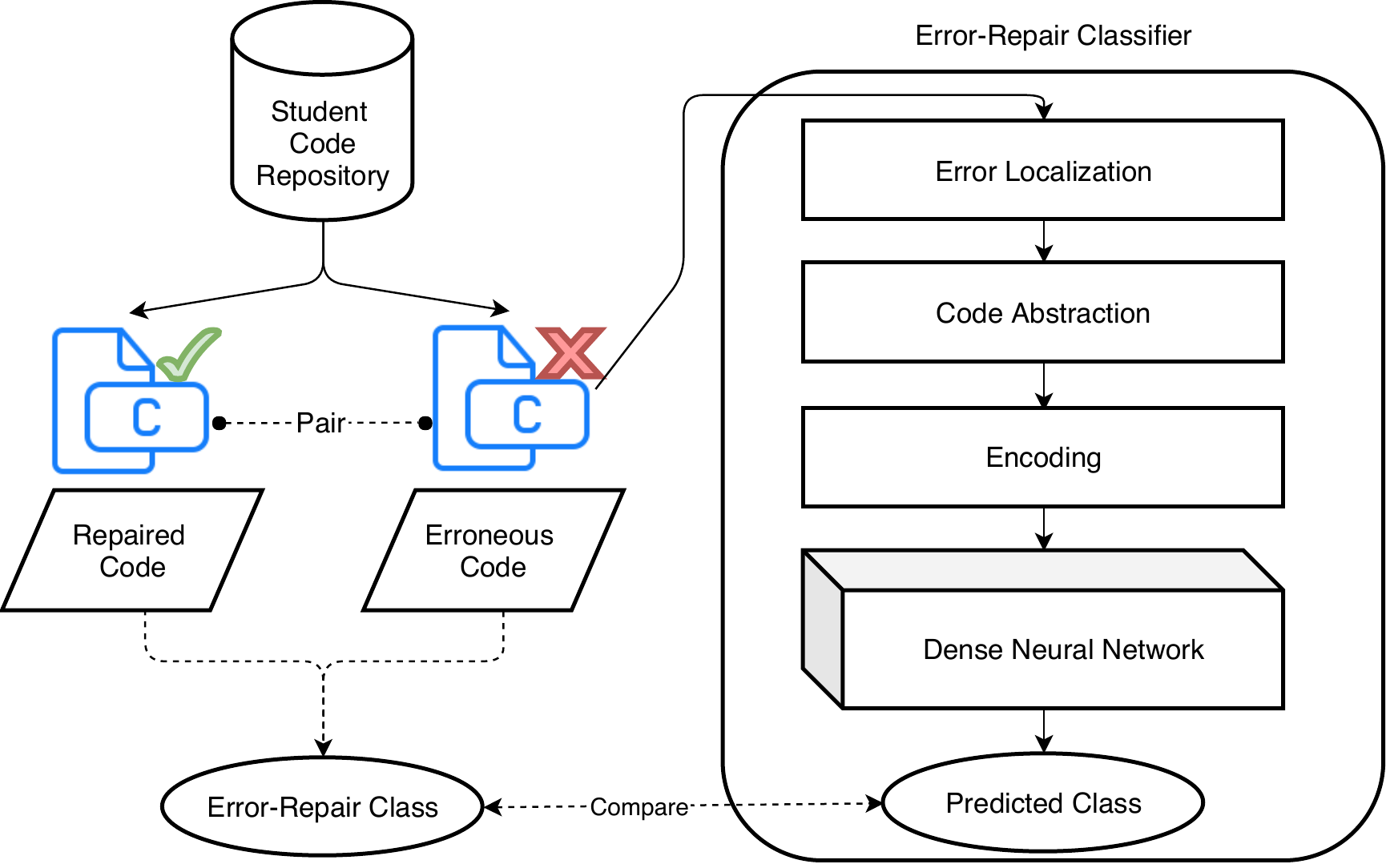}
            \caption{Training Phase}
            \label{fig:eg_systemTrain}
        \end{subfigure} \hspace{0.5em} 

        \begin{subfigure}[t]{.29\linewidth}   
            \includegraphics[width=\linewidth, keepaspectratio]{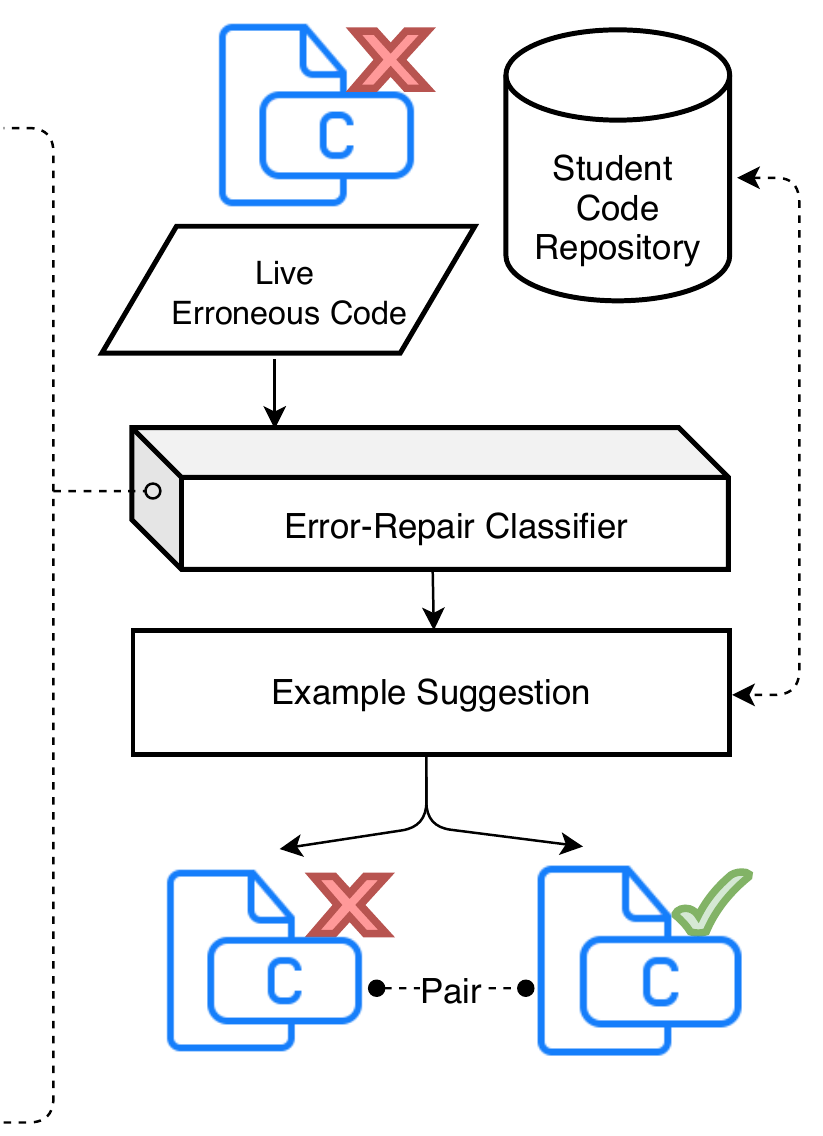}
            \caption{Live Deployment Phase}
            \label{fig:eg_systemTest}
        \end{subfigure}\hfill     
    }

\caption{\tegcer system workflow. The error-repair classifier is trained on previous student code repository during the training phase. In live deployment phase, the trained classifier is used to predict and fetch relevant examples for live erroneous codes.}
\label{fig:eg_system} 
\end{figure*}

\subsection*{Our Contributions}

Our main contribution is the theory and implementation of  \tegcer\footnote{\label{github}\url{https://github.com/umairzahmed/tegcer}}, a tool to automatically suggest relevant programming examples as feedback to students, without requiring any human assistance.  To the best of our knowledge, \tegcer is the first automated tool of this kind.  In the extremely challenging task of predicting a single correct label for buggy program from 212 unique error-repair classes, \tegcer achieves very high accuracy of $87\%$ on its first prediction (\acci{1}).

From a dataset of 15,000+ buggy programs, we identified 6,700+ unique compilation error-groups ($EG$s) made by novice students, and 200+ different class of error-repairs ($C$s) performed by them to fix the errors. This dataset is released in public domain to help further research\footnotemark[1]{}.

Finally, we also report results from a large-scale empirical evaluation of our system, comparing its ability to assist students with that of human teaching assistants, across errors of greatly varying complexity.
\section{\tegcer System Overview}
\label{sec:eg_tool}

In this section, we present \tegcer in its entirety. Figures~\ref{fig:eg_systemTrain} and \ref{fig:eg_systemTest} give an overview of the training and live deployment phases respectively.


From a course offering of CS1 at IIT Kanpur, a large public university, we obtained a dataset of $15,000+$ erroneous student programs which fail to compile, along with their corresponding repaired code. During the training phase, we train an error-repair classifier on this dataset, to predict the type of error-repair class it belongs to. As mentioned in Figure~\ref{fig:eg_systemTrain}, this classifier involves a four-phased methodology as follows.

\subsection{Error Localization}
Given an erroneous code, \tegcer locates the line(s) where repair must be performed, by relying on the exact line number  reported by the compiler. 
This is a departure from existing repair tools, such as \tracer~\cite{ahmed2018compilation} and DeepFix~\cite{gupta2017deepfix}, which employ a ``search, repair \& test'' strategy to achieve localization accuracy of $86\%$. Such a strategy is not applicable for example generation, due to its very nature of being subjective. \tegcer achieves localization accuracy of $80\%$ on the same dataset, by focusing on only compiler reported lines. 
\subsection{Code Abstraction}
As a second step, \tegcer abstracts the compiler reported lines, by replacing program specific tokens such as variable names, function names, literals, etc, with their generic types. These types are inferred using LLVM~\cite{lattner2004llvm}, a standard static analysis tool. This abstraction module is largely motivated by the one employed in \tracer~\cite{ahmed2018compilation}.
This stage greatly reduces the load on neural network classifier (described in Section~\ref{neuralnetwork}), and helps it generalize better, by performing implicit vocabulary compression.

\subsection{Encoding}
The abstract tokens are then encoded into a binary representation, with few additional ``features'' added, before passing them to our neural network. This technique is further elaborated in Section~\ref{subsec:eg_dataPrep}.

\subsection{Dense Neural Network} \label{neuralnetwork}
Finally, we train a dense neural network classifier, which takes the encoded erroneous line, and learns to predict the error-repair class that it belongs to. The error-repair class captures the type of mistake made by student, and its desired repair. The concept of error-repair class and our neural network setup are explained in Sections~\ref{sec:eg_class} and~\ref{sec:eg_classifier} respectively.


\subsection*{Live Deployment}
Figure~\ref{fig:eg_systemTest} shows the workflow during live deployment phase. Given a new erroneous code encountered in the live setting, the previously trained classifier is used to predict the error-repair class for each  erroneous line. The old student-code repository is then queried to fetch all examples that belong to the exact same predicted class. The repaired code pair of these examples is then suggested back to student as feedback, in the decreasing order of their frequency of occurrence.

\section{Student Code Repository}
\label{sec:eg_data}

The student code repository consists of code attempts made by students, during the 2015--2016 fall semester course offering of Introductory to C Programming (CS1) at IIT Kanpur, a large public university. This course has been credited by $400+$ first year undergraduate students, who attempted $40+$ different programming assignments as part of course requirement. These assignments were completed on a custom web-browser based IDE, which records all intermediate code attempts. This dataset has previously been used by multiple state-of-the-art repair tools in literature~\cite{clara,gupta2017deepfix,ahmed2018compilation,gupta2018deep}.

From the logs for the entire semester, we found a total of $23,275$ program pairs such that (i) the student program failed to compile and (ii) the same student edited a single-line in the buggy program to repair it. We use only these single-line edit programs during our training phase. During the live deployment phase, programs with errors on multiple-lines are treated as multiple instances of single-line errors. That is, during the live deployment phase, \tegcer can handle programs with errors on multiple-lines by suggesting examples for each compiler reported erroneous line.

\begin{figure}[tbp] 
    \centering
    \mbox{ 
        \centering
        \hspace{0.2em}                
        \begin{subfigure}[t]{.55\linewidth}
            \centering \small
\begin{lstlisting}[]
void main(){
    int i=0;    
    if($\text{\colorbox{red!30}{\!0 = i\!}}$)
        i++;
}
\end{lstlisting}
            
            \hspace{-1.7em}
            \begin{tabular}{l@{\ \ }l@{\ \ }p{3.5cm}}
                & & \textbf{Compiler Message}\\ 
                $3$ & $E_{10}$ & {expression is not assignable} \\
            \end{tabular}

            \caption{Erroneous program}
            \label{fig:eg_rc8bad}
        \end{subfigure}	

        \begin{subfigure}[t]{.4\linewidth}
            \centering \small
\begin{lstlisting}[]
void main(){
    int i=0;    
    if($\text{\colorbox{green!30}{\!0 == i\!}}$)
        i++;
}
\end{lstlisting}
            
            \begin{tabular}{l@{\ }l@{\ }p{3cm}}
                && \textbf{Repair Tokens}\\ 
                && \{\code{+==}, \code{-=}\} \\
            \end{tabular}
        
            \caption{Repaired program}
            \label{fig:eg_rc8good}
        \end{subfigure}	\hfill
    }
\caption{Erroneous-Repaired code pair
         \{$E_{10}$ \code{+==\quad-=}\}}
\label{fig:eg_rc8}
\end{figure}

\subsection{Error Types}

\begin{table}[tbp] 
    \centering
    \begin{tabular}{l l | l l}
        \toprule            
        \textbf{Error ID} & \textbf{Message} & \textbf{$\dottedsquare_1$} & \textbf{$\dottedsquare_2$} \\ 
        \midrule
        $E_1$ & {Expected $\dottedsquare_1$} & ) & \\
        $E_2$ & {Expected $\dottedsquare_1$ after expression} & ; & \\
        $E_3$ & {Use of undeclared identifier $\dottedsquare_1$} & sum & \\
        $E_4$ & {Expected expression} & & \\
        $E_5$ & {Expected identifier or $\dottedsquare_1$} & ( & \\
        $E_6$ & {Extraneous closing brace ($\dottedsquare_1$)} & \} & \\
        $E_7$ & {Expected $\dottedsquare_1$ in $\dottedsquare_2$ statement} & ; & for\\
        $E_8$ & {Expected $\dottedsquare_1$ at end of decl.} & ; & \\
        $E_9$ & Invalid operands to binary & {int *} & {int} \\
              & expression ($\dottedsquare_1$ and $\dottedsquare_2$) & & \\
        $E_{10}$ & {Expression is not assignable} & & \\
        \bottomrule
    \end{tabular}
\caption[Top frequent compilation errors (Es)]{The top--10 frequent individual compilation errors (Es), listed in the decreasing order of frequency.}
\label{tab:eg_top10ce}
\end{table}

The $23,275$ erroneous code attempts trigger $250+$ unique kinds of compilation error messages. Table~\ref{tab:eg_top10ce} lists the top-5 frequent error messages returned by Clang compiler~\cite{lattner2004llvm}, a popular compiler for C programming language. The messages are generalized by replacing any program specific tokens (demarked by Clang within single/double quotes) with $\dottedsquare$. One example substitution for each $\dottedsquare$ is also provided in the table. 

Figure~\ref{fig:eg_rc8bad} presents an erroneous code example that suffers from compilation error $E_{10}$ (expression is not assignable), due to incorrect usage of assignment operator ``\code{=}'', in place of equality operator ``\code{==}''.

\subsection{Error Groups}

\begin{table}[tbp] 
    \centering
    \begin{tabular}{l | l l l l l }
        \toprule
        \textbf{Error Group} & $EG_1$ & $EG_2$ & $EG_3$ & $EG_4$ & $EG_5$  \\ 
        \textbf{Error} & $E_3$ & $E_4$ & $E_1$ & $E_2$ & $E_{35}$ \\
        
        \midrule
        
        \textbf{Error Group} & $EG_6$ & $EG_7$ & $EG_8$ & $EG_9$ & $EG_{10}$ \\
        \textbf{Error} & {$E_3 \land E_4$} & $E_5 \land E_6$ & $E_{9}$ & $E_1 \land E_4$ & $E_5$ \\
              
        \bottomrule
    \end{tabular}
    \caption{Top--10 frequent compilation error-groups (EG).}
\label{tab:eg_topGroups}
\end{table}

\begin{figure}[t!] 
    \centering
    \includegraphics[width=.7\linewidth, keepaspectratio]{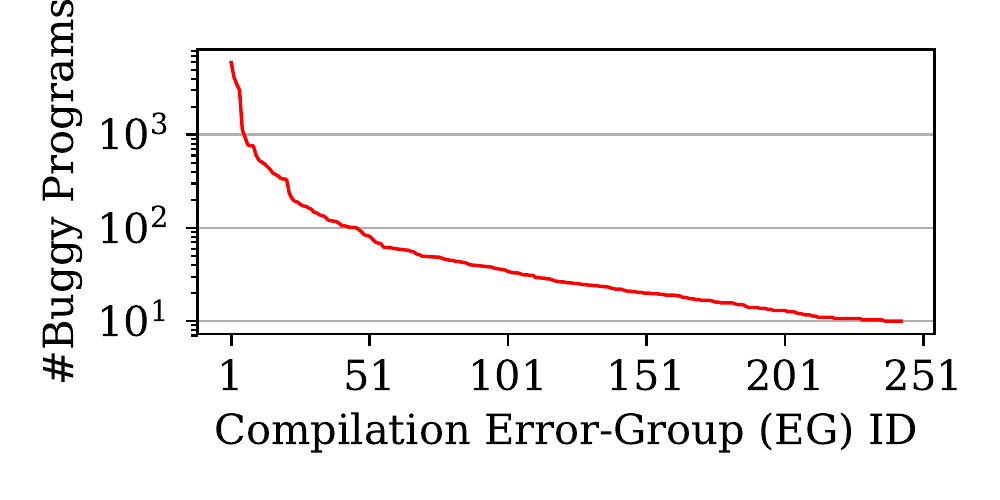}
    \caption{Frequency distribution of compilation error-groups ($EG$s). The plot depicts the \#program (y-axis, in $\log_{10}$ scale) attempts by students, that failed to compile due to presence of $EG$ (x-axis).}
\label{fig:eg_errHist}
\end{figure}

A buggy program can contain one or more compilation errors, a collection of which is called compilation error group (abbreviated as \textbf{EG}). In the dataset of the course offering, the $250+$ unique individual compilation errors ($E_i$) combine to form $6,700+$ unique error groups ($EG$s). 

This grouping is performed since the bug in the program (and hence its fix) is better characterized by the set of errors occurring together, as opposed to each individual error considered independently. For example, error-group $EG_7 = E_5 \land E_6$ typically occurs due to mismatched braces, whose repair involves inserting/deleting ``\code{\{}'' or ``\code{\}}''. Where as, the error group $EG_{10}$ containing single error $E_5$ typically occurs when the program contains a spurious comma operator ``\code{,}''. For example, the line ``\code{int i,j,;}'' suffers from $EG_{10} = E_5$ error group/type.

Table~\ref{tab:eg_topGroups} lists the top frequently observed $EG$s. $EG_1$ (resp $EG_{10}$) is the top-frequent (resp $10^{th}$ frequent) compilation error-group, encountered by close to $6,000$ (resp $600$) programs compiled by students. 

As observed in Fig.~\ref{fig:eg_errHist}, the frequency plot of the set of errors ($EG$s) novice programmers make in our course offering follows a heavy tailed distribution. In other words, there is a sharp decrease in the number of programs affected by an error group ($EG$s), as we proceed through a list of $EG$s sorted by their frequency count. Similar observation has been made by others, on their student error datasets~\cite{ahmed2018compilation,denny2012all,hartmann2010helpmeout}. In the entire course offering where students made $23,000+$ failure code attempts, the top-8 frequent $EG$s accounted for more than $50\%$ of all the student programs failing to compile. Out of the $6700+$ different $EG$s, only the top-$240$ ones repeat in $10$ or more different programs, for the entire run of the course offering.

\subsection{Repair Tokens}
For every erroneous code in the dataset, we have a corresponding repaired code pair. This repair is performed by the same student, by inserting and deleting program tokens until all compilation errors are resolved. We treat replacements as a combination of insertion and deletion. 

After performing error-localization and code-abstraction on $23,000+$ erroneous-repaired code pairs, the difference between these two abstract lines form the set of repair tokens ($R$s). The top-5 repair tokens inserted/deleted by students are ``\code{INVALID}'', ``\code{;}'', ``\code{,}'', ``\code{)}'' and ``\code{INT}''. The ``\code{INVALID}'' abstract token represents an undeclared variable/function and ``\code{INT}'' refers to integer datatype variables.

The repaired code example in Figure~\ref{fig:eg_rc8good} deletes assignment operator ``\code{=}'', and inserts an equality operator ``\code{==}''. Hence its set of repair tokens are \{\code{+==}, \code{-=}\}.

\subsection{Error Repair Class}
\label{sec:eg_class}

\begin{table}[tbp] 
    \centering
    \begin{tabular}{l | l | r}
        \toprule
            \textbf{Class ID} & \multicolumn{1}{l|}{\textbf{Error Repair Class}} & \textbf{\#Programs} \\
        \midrule
            $C_1$ & $E_2\quad$ \code{+;} & $3,888$\\
            $C_2$ & $E_3\quad$ \code{+INT\quad-INVALID} & $731$ \\
            $C_3$ & $E_1\quad$ \code{+\}} & $519$ \\
            $C_4$ & $E_1\quad$ \code{+)} & $475$ \\
            $C_5$ & $E_5 \land E_6\quad$ \code{-\}} & $418$ \\
            $C_6$ & $E_8\quad$ \code{+;} & $404$ \\
            $C_7$ & $E_1\quad$ \code{+,} & $389$ \\
            $C_8$ & $E_{10}\quad$ \code{+==\quad-=} & $309$ \\
            $C_9$ & $E_1\quad$ \code{+;} & $305$ \\
            $C_{10}$ & $E_7\quad$ \code{+;\quad-,} & $299$ \\

            \multicolumn{1}{c}{\ldots} & \multicolumn{1}{c}{} & \multicolumn{1}{c}{} \\
            $C_{211}$ & $E_{47}\quad$ \code{-,\quad-INT} & $10$ \\ 
            $C_{212}$ & $E_1 \land E_5 \land E_{47}\quad$ \code{-;} & $10$ \\ 
        \midrule
            ALL & \multicolumn{1}{c|}{--} & $15,579$ \\   
        \bottomrule
    \end{tabular}
    \caption{Top frequent error repair classes ($C$s) from single-line edit dataset. The class is a combination of error types ($E$s) and repair tokens ($R$s). A total of $212$ classes contain $10$ or more buggy programs.}
\label{tab:eg_topRC}
\end{table}

\begin{table}[t!] 
    \centering 
    \begin{tabular}{l @{} r | l | l | r}
        \toprule
            \textbf{$EG$} & \textbf{\#$C$s} & \textbf{$C$} & \multicolumn{1}{l|}{\textbf{Error Repair Class ($C$)}} & \textbf{\#Programs} \\
        \midrule
            \multirow{3}{*}{$EG_1$} & \multirow{3}{*}{24}  
                    &  $C_2$ & $E_3$ \code{+INT} \code{-INVALID} & 731 \\
                    && $C_{11}$ & $E_3$ \code{+ARRAY} \code{-INVALID}  & 286 \\
                    && $C_{16}$ & $E_3$ \code{+LITERAL\_INT} \code{-INVALID}  & 150 \\

        \midrule
            \multirow{3}{*}{$EG_2$} & \multirow{3}{*}{38}    
                    &  $C_{15}$ & $E_4$ \code{+INT} & 190 \\
                    && $C_{17}$ & $E_4$ \code{+LITERAL\_INT} & 145 \\
                    && $C_{18}$ & $E_4$ \code{-[\quad-]} & 138 \\

        \midrule
            \multirow{3}{*}{$EG_3$} & \multirow{3}{*}{28}     
                    &  $C_3$ & $E_1$ \code{+\}} & 519 \\
                    && $C_4$ & $E_1$ \code{+)}  & 475 \\
                    && $C_7$ & $E_1$ \code{+,}  & 389 \\
        
        \midrule
            \multirow{3}{*}{$EG_4$} & \multirow{3}{*}{5}    
                    &  $C_1$ & $E_2$ \code{+;} & 3,888 \\ 
                    && $C_{58}$ & $E_2$ \code{+if} & 52 \\
                    && $C_{79}$ & $E_2$ \code{+;\quad-:} & 36 \\    

        \midrule
            \multirow{2}{*}{$EG_7$} & \multirow{2}{*}{2}  
                    &  $C_5$ & $E_5 \land E_6$ \code{-\}} & 418 \\ 
                    && $C_{78}$ & $E_5 \land E_6$ \code{+\{} & 41 \\

        \midrule
            \multirow{3}{*}{$EG_{10}$} & \multirow{3}{*}{3} 
                    &  $C_{22}$ & $E_5$ \code{-,} & 109 \\ 
                    && $C_{205}$ & $E_5$ \code{-;} & 10 \\
                    && $C_{206}$ & $E_5$ \code{-int} & 10 \\
        
        \bottomrule
    \end{tabular}
    \caption[Repair Classes (RCs) per Error-Group (EG)]{Top-3 Repair Classes ($C$s), for few of the frequent error-group ($EG$s). The second column (\#$C$s) denotes the total number of different repairs possible for a given $EG$, with the third and fourth column listing the top-3 frequent class ID ($C$-ID) and repair class ($C$) respectively. Finally, the number of buggy programs (\#Programs) belonging to each $C$ is provided in the last column.}
\label{tab:eg_RC_EG}
\end{table}

Given a buggy source program that suffers from compilation errors ($E$s) which require a set of repair tokens ($R$s) to fix, its \textit{error-repair class} ($C$) is defined as the merged set of errors and repairs \{$E$s $R$s\}. For example, the erroneous-repaired code pair in Figure~\ref{fig:eg_rc8} belongs to $C_8$ \{$E_{10}$ \code{+==\quad-=}\}, the $8^{th}$ most frequently occurring error-repair class. 

We determine the error-repair class of the $23,275$ erroneous-repaired code pairs in our dataset. Table~\ref{tab:eg_topRC} lists the error-repair classes ($C$s) sorted in decreasing order of frequency, along with the number of buggy programs belonging to each class. A total of $212$ classes were found containing $10$ or more buggy programs, and total of $15,579$ programs belong to one of these $212$ classes. Only these $15,579$ buggy programs and their $212$ classes then form the training dataset for our deep classifier. While the remaining classes are unused due to lack of sufficient training examples.

As seen from Table~\ref{tab:eg_topRC}, $C_1$ \{$E_2$ \code{+;} \} is the top most frequent error repair class, containing $3,888$ buggy programs which encounter compilation error $E_2$ (expected $\dottedsquare_1$ after expression) and require one or more insertion of semi-colon ``\code{+;}'' to fix it. While $C_{212}$ \{$E_1 \land E_5 \land E_{47}$ \code{-;} \} is one of the least frequent class containing $10$ buggy programs, which encounter compilation errors $E_1 \land E_5 \land E_{47}$ and require one or more deletion of semi-colon ``\code{-;}'' to fix it. Additional examples of error-repair classes are presented later in Section~\ref{sec:eg_egSuggestion}.

Table~\ref{tab:eg_RC_EG} presents a different arrangement of error-repair classes, grouped based on the error-group ($EG$). As seen from this table, a wide variety of repairs were undertaken by our students, for the same set of compiler reported error messages ($E$s). Consider the last 3 row of Table~\ref{tab:eg_RC_EG} for error-group $EG_{10}$, which consists of single compiler error $E_5$ (Expected identifier or $\dottedsquare_1$). Students in our course offerring typically resolved this error in 3 unique ways; either by deleting commas (\code{-,}), deleting semi-colons (\code{-;}), or by deleting the keyword \code{int}. The choice of repair is dictated by program context, which our error-repair class attempts to capture. Maintaining a trivial list of programs per compilation error, which can then be retrieved and suggested as examples, would not capture the relevant repair.

\section{Error-Repair Classifier}
\label{sec:eg_classifier}
In this section, we present our deep neural network setup which, given an erroneous abstract line, predicts the relevant error-repair class that it belongs to. 

\subsection{Data Encoding}
\label{subsec:eg_dataPrep}

The labelled dataset of $15,579$ erroneous programs is split in the ratio of $70\%:10\%:20\%$ for training, validation and testing purpose respectively. Before the abstracted erroneous line is supplied to neural network for learning,  we pre-process it to help the deep network generalize better. 

\subsubsection{Input Tokens}

\begin{table}[t!] 
    \centering
    \begin{tabular}{l| l c l}
        \toprule
        & \textbf{Erroneous} & & \textbf{Repaired} \\ 
        \midrule
        \textbf{Line} & \code{b=xyz;} & 
            $\longrightarrow$ & \code{b=a;} \\
        \textbf{Abstraction} & \code{INT=INVALID;} & 
            $\longrightarrow$ & \code{INT=INT;} \\ \midrule

        \textbf{Repair} & & \code{+INT -INVALID} & \\
    \end{tabular}    
    \caption{Example erroneous-repaired line pair, requiring replacement of ``\code{INVALID}'' token with ``\code{INT}'' token. The compiler reports error $E_3$ (use of undeclared identifier $\dottedsquare_1$).}
\label{tab:eg_repairRepl} 
\end{table}

For each erroneous abstract line, the input to neural network consists of unigram tokens, bigram tokens, and compilation errors ($E$s). For example, consider the erroneous line shown in Table~\ref{tab:eg_repairRepl}. The sequence of input tokens passed to deep network are: \{\code{<ERR>}, \code{$E_3$}, \code{<UNI>}, \code{INT}, \code{=}, \code{INVALID}, \code{;}, \code{<BI>}, \code{INT\_=}, \code{=\_INVALID}, \code{INVALID\_;}, \code{<EOS>}\}.

In our dataset of $15,579$ source lines, the input vocabulary size is observed to be $1,756$. That is, a total of $1,756$ unique input tokens (combination of unigrams, bigrams and $E$s) exist in our dataset. 

\subsubsection{Input Encoding}
Each input token is then vectorized using \textit{tokenizer\footnote{\url{https://keras.io/preprocessing/text/\#tokenizer}}}, a text pre-processor provided by the \textit{Keras} deep learning library~\cite{chollet2015keras}. In this encoding, the input tokens are represented using one-hot encoding---a binary representation indicating presence or absence of token. 

More involved encoding techniques such as frequency or tf-idf~\cite{rajaraman_ullman_2011} gave poor results, due to the extreme paucity of data for most classes; More than $30\%$ of the $212$ classes have just $10$ erroneous examples to train and test from. 

\subsubsection{Output Encoding}
Similar to the input encoding, output error-repair class labels need to be encoded as well, before neural network learns to predict them. The $212$ error-repair class labels are represented with one-hot binary encoding, using the \textit{to\_categorical\footnote{\url{https://keras.io/utils/\#to_categorical}}} utility provided by \textit{Keras}.

\subsection{Dense Neural Network}
\label{subsec:eg_neuralNet}

Allamanis \etal~\cite{allamanis2018survey} present an extensive survey of existing deep network setups used in literature, to represent computer programs for various tasks. For our error-repair classification task, we experimented with some of the popular complex deep networks, such as Long Short Term Memory (LSTM) models with embeddings~\cite{hochreiter1997long} and Convoluted Neural Network (CNN) models with max-pooling~\cite{krizhevsky2012imagenetCNN}. These complex models need to train a large number of parameters, in the order of millions of different weights, capturing patterns in the entire sequence of input tokens, for which massive amounts of training data is required, typically in the order of thousands per class. 

However, in our labelled dataset of $15,000$+ programs tagged with $212$ labels, only the largest class has more than $1,000$ programs. On the other hand, $67$ of these classes ($31\%$ of total $212$) have just $10$ examples each, for both training and testing the classifier. Due to which complex neural networks fail to generalize on our dataset of student errors, recording prediction accuracy in the range of $30-40\%$.

\subsubsection{Neural Network Layers}

\begin{figure}[tbp] 
    \centering
    \includegraphics[width=\linewidth, keepaspectratio]{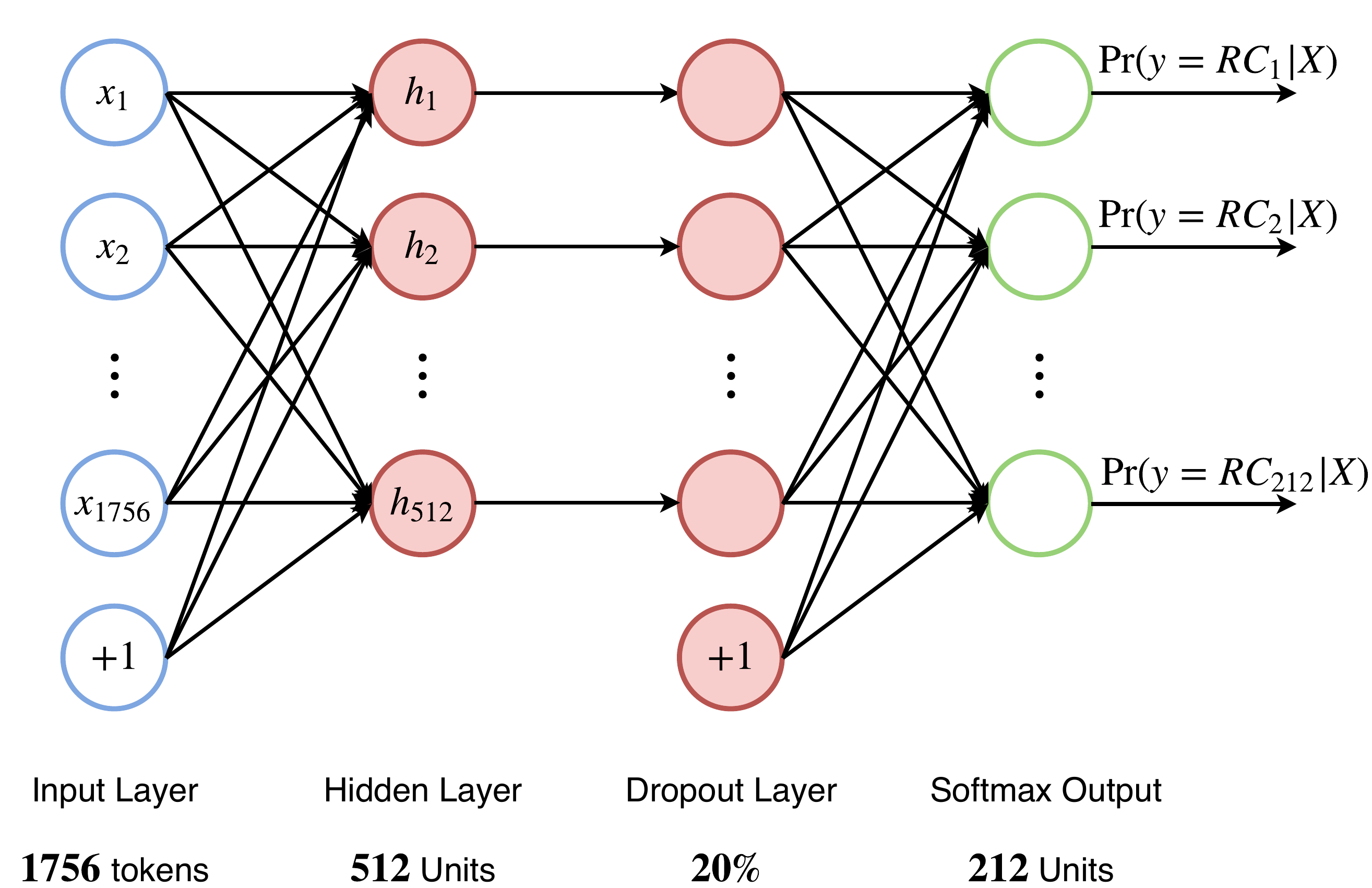}
    \caption[Dense neural network classifier]{Dense Neural Network of $512$ hidden units with $20\%$ dropout, used to classify $1,756$ input tokens with $212$ error-repair class labels.}
\label{fig:eg_denseNet}
\end{figure}

To overcome the difficulties mentioned above, we turn towards simpler deep network model for our class predictions. In particular, we found that a dense neural network with single hidden layer is best suited for our task. A dense network is a fully connected network, where each neuron is connected to all the neurons of previous layer. 

Figure~\ref{fig:eg_denseNet} describes the network arrangement of our dense classifier. The first layer is an input layer, which consists of $1,756$ binary encoded input source tokens. The second layer is the single hidden layer of $512$ units, densely connected with the previous input layer. This is followed by a dropout layer~\cite{srivastava2014dropout}, which randomly drops $20\%$ of the input neurons to $0$ during the training stage, to avoid overfitting the model on training dataset. Finally, our last layer is an output layer containing $212$ units (one for each class). This output layer is densely connected with the previous hidden layer as well.

\subsubsection{Parameters}
\textit{Keras}~\cite{chollet2015keras} framework was used to build our dense neural network classifier, with \textit{TensorFlow}~\cite{abadi2016tensorflow} backend. The following parameters, including the hidden layer size, were selected based on performance of classifier on the $10\%$ validation set.

We used Rectified Linear Unit (ReLU), a non-linear activation function, between the hidden layers. While a softmax activation function is used in the last output layer, to predict the probabilities of all $212$ classes, given the set of input tokens. 

Stochastic gradient descent algorithm was used to learn the weights between layers, by minimizing a categorical cross-entropy loss function over the $70\%$ training data set. Also, Adam~\cite{kingma2014adam}, an adaptive moment estimator function, was chosen to optimize the learning process, leading to faster and stable convergence. 

The peak accuracy of our model on the $10\%$ validation data set was obtained after training for a short number of $6$ epochs. Beyond this, the model begins to heavily overfit on training dataset, due to the small number of examples available for most classes. The training phase typically lasts 1-2 minutes, while predicting a class during the testing phase requires few milliseconds.

\subsection{Accuracy}

\begin{table}[tbp] 
    \centering
    \begin{tabular}{l r r r}
        Classifier & \acci{1} & \acci{3} & \acci{5} \\
    \toprule
        Dense Neural Network & $87.06\%$ & $97.68\%$ & $98.81\%$ \\ 
    \end{tabular}
\caption[Dense neural network classifier accuracies]{\acci{k} accuracy for  Dense Neural Network classifier}
\label{tab:eg_classifierAcc}
\end{table}
            
\begin{table*}[ht!] 
    \centering
    \begin{tabular}{l | ll | rr rr | llr}
        \toprule
            \textbf{$C_{ID}$} & \multicolumn{2}{l|}{\textbf{Error-Repair Class ($C$)}} & 
                \textbf{\#Train} & \textbf{\#Test} & \textbf{Precision} & \textbf{Recall} &
                \multicolumn{3}{c}{\textbf{Top Incorrect Prediction}}\\
            & & & & & & & \multicolumn{2}{l}{\textbf{Incorrect Class}} & \textbf{\#Test} \\
        \midrule
            $C_1$ & $E_2$ & \code{+;} & 3,110 & 778 & 0.99 & 0.99 
                & $E_2$ & \code{+if} & 4 \\
            $C_2$ & $E_3$ & \code{+INT\quad-INVALID} & 585 & 146 & 0.79 & 0.79 
                & $E_3$ & \code{+LITERAL\_INT\quad-INVALID} & 7 \\
            $C_3$ & $E_1$ & \code{+\}} & 415 & 104 & 0.91 & 0.95 
                & $E_1$ & \code{-\{} & 2 \\
            $C_4$ & $E_1$ & \code{+)} & 380 & 95 & 0.77 & 0.84 
                & $E_1$ & \code{-(} & 10 \\
            $C_5$ & $E_5 \land E_6$ & \code{-\}} & 334 & 84 & 1.00 & 1.00 
                & \multicolumn{3}{c}{--} \\
            &&&&&\\

            $C_6$ & $E_8$ & \code{+;} & 323 & 81 & 0.97 & 0.95 
                & $E_8$ & \code{+; -,} & 4 \\
            $C_7$ & $E_1$ & \code{+,} & 311 & 78 & 1.00 & 0.96 
                & $E_1$ & \code{+\}} & 3 \\
            $C_8$ & $E_{10}$ & \code{+==\quad-=} & 247 & 62 & 0.97 & 0.98 
                & $E_{10}$ & \code{+(\quad+)} & 1 \\
            $C_9$ & $E_1$ & \code{+;} & 244 & 61 & 0.97 & 0.98 
                & $E_1$ & \code{+)} & 1 \\
            $C_{10}$ & $E_7$ & \code{+;\quad-,} & 239 & 60 & 1.00 & 0.97  
                & $E_7$ & \code{+;} & 2\\

            \multicolumn{3}{c}{\ldots} \\
            $C_{211}$ & $E_{47}$ & \code{-,\quad-INT} & 8 & 2 & 1.00 & 1.00 
                & \multicolumn{3}{c}{--} \\ 
            $C_{212}$ & $E_1 \land E_5 \land E_{47}$ & \code{-;} & 8 & 2 & 1.00 & 1.00
                & \multicolumn{3}{c}{--} \\ 
        \bottomrule
    \end{tabular}
    \caption{Precision and Recall Scores of Dense Neural Network on top frequent classes. The columns \#Train and \#Test indicates the total number of source lines used for training/validation and testing of the classifier, respectively. For each actual class $C$, \textit{Top Incorrect Prediction} column lists the top incorrectly predicted class, along with the number of test-cases in which this mis-classification occurred.}
    \label{tab:eg_precRecallRC}
\end{table*}

We report on the overall accuracy of the classifier, as well as precision-recall scores of individual classes, on our held out $20\%$ test dataset. The latter analysis is necessitated due to the highly skewed distribution of classes in our dataset, where the top-10 classes, out of the $212$ unique ones, account for $50\%$+ of the total test-cases. 

\subsubsection{Overall Accuracy}

We measure the overall performance using a \acci{k} metric, which denotes the percentage of test-cases where the top-\kac class predictions contains the actual class. The classifier returns a probability score for each class on being given a buggy line, $Pr(y=C_j|X)$. Which can be sorted in descending order to select the top-\kac results as predicted classes.

Table~\ref{tab:eg_classifierAcc} reports on the accuracy of dense network in predicting classes on $20\%$ held out test-set of $3,098$ buggy source lines, for various values of \kac. When we consider only the top prediction (\acci{1}) by the dense classifier, the predicted class is exactly same as the actual class in $87.06\%$ of the cases. If we are to sample the top-3 predictions instead, then the actual class is present in one of these three predictions for  $97.68\%$ of the cases. In other words, for majority of the test cases ($3026$ out of $3098$), dense neural network predicts the correct class in its top-3. 


\subsubsection{Top Frequent Classes}
\newcommand{\bk}{{j}}

Table~\ref{tab:eg_precRecallRC} lists the precision and recall scores of our dense neural network for the top-5 frequent classes ($C$s). The classifier enjoys high precision/recall scores, of more than $0.9$, across most of the listed classes. Further, even on the extremely rare classes of $C_{211}$ and $C_{212}$, that offer just $8$ buggy examples to train and validate from, the neural network is able to generalize successfully and achieve $100\%$ precision/recall score. 

Two of the listed classes, $C_2$ and $C_4$, suffer from (relatively) weaker recall scores of $0.79$ and $0.84$, respectively. In other words, out of the $146$ (resp. $95$) test cases having label $C_2$ (resp. $C_4$), the classifier is able to correctly predict $79\%$ (resp. $84\%$) of them. The last column, which lists the top most incorrect prediction made by classifier, are both valid classes.

From Table~\ref{tab:eg_precRecallRC}, the class $C_2$ \{$E_3$ \code{+INT\quad-INVALID}\} is most confused with \{$E_3$ \code{+LITERAL\_INT\quad-INVALID}\}. This is intuitive since, in C programming language, a variable is often interchangeable with a literal of the same type. Similarly, the class $C_4$ \{$E_1$ \code{+)}\} is incorrectly predicted as class \{$E_1$ \code{-(}\} $10$ different times, since the repair for incorrectly balanced parenthesis could be to either add more closing parenthesis, or remove few of the existing open ones. While choice of the repair is dictated by logical correctness, either of these two can be used to fix the compilation error. 

This suggests a larger issue, where an erroneous line can belong to multiple classes, since it can be repaired in multiple different ways. Since our erroneous-repaired dataset captures only one form of repair performed by actual student, we are forced to treat the problem as multi-class, instead of multi-label. This limitation accounts for the lower precision/recall scores observed in some of the classes. 


\subsubsection{Individual Classes}

\begin{figure}[tbp] 
    \centering
    \mbox{ 
        \centering
        \begin{subfigure}[t]{.47\linewidth} 
            \includegraphics[width=\linewidth, keepaspectratio]{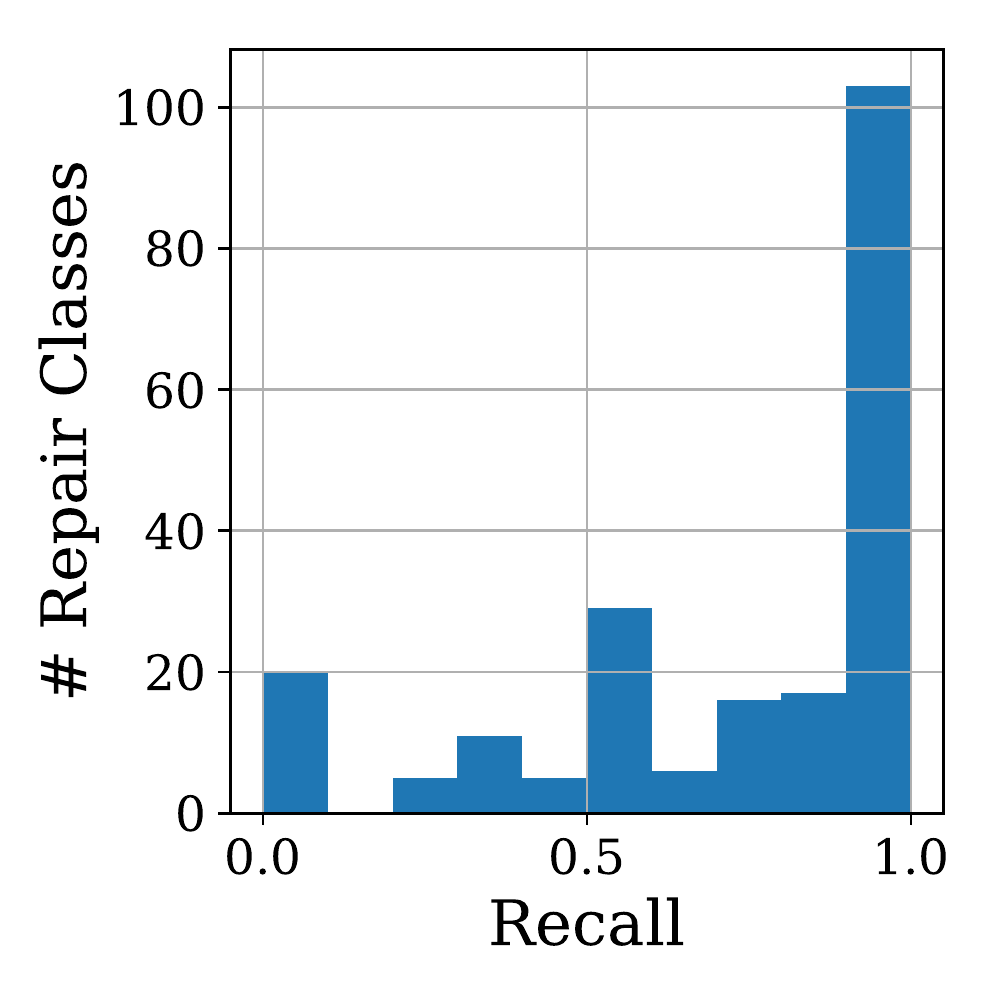}
            \caption{Histogram of $212$ classes' recall scores}
            \label{fig:eg_recall_classC}
        \end{subfigure} 

        \begin{subfigure}[t]{.47\linewidth}   
            \includegraphics[width=\linewidth, keepaspectratio]{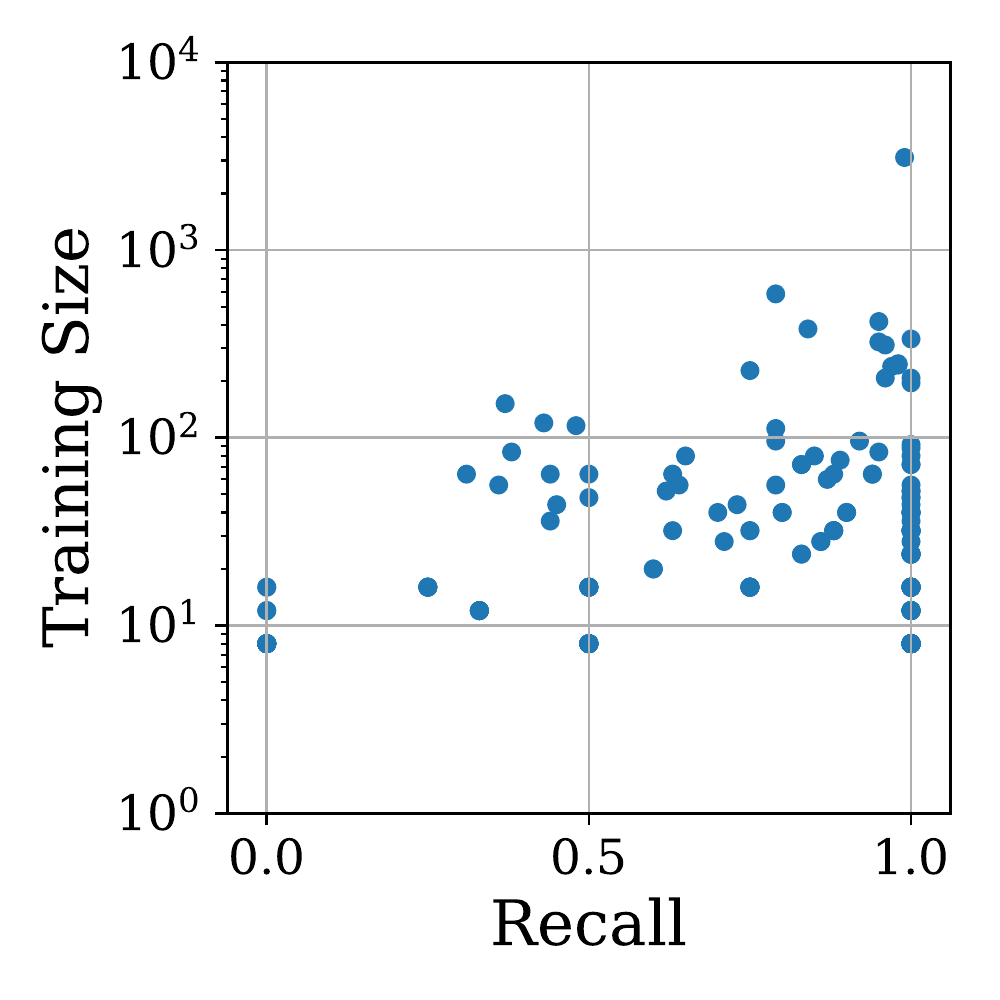}
            \caption{Effect of $212$ classes' training size on recall scores}
            \label{fig:eg_recall_size}
        \end{subfigure}\hfill     
    }

\caption[Dense neural network recall scores]{Dense neural network recall scores for $212$ classes}
\label{fig:eg_recall} 
\end{figure}

In Figure~\ref{fig:eg_recall}, we analyze the recall scores across all $212$ classes. From Figure~\ref{fig:eg_recall_classC} it is seen that, our dense neural classifier achieves high accuracy, across majority of the classes. The recall score is $>=0.5$ on $80\%+$ ($171/212$) of classes, and $>=0.9$ for more than $100$ classes.

Figure~\ref{fig:eg_recall_size} then presents the recall scores of all $212$ classes, plotted against their training sample size. While the classifier achieves $100\%$ recall score on classes across multiple training sizes, including those with just $8$ training examples, the recall scores drop to $<=0.5$ (resp. $0.0$), only when the training size is below $200$ (resp. $20$). In other words, having more training examples for small sized classes could help improve their prediction scores.

\begin{table*}[htbp] 
    \centering    
    \begin{tabular}{l l ll c}
        \toprule
        \# & {Erroneous-Line and } & {Predicted}       & {\tegcer's Top Example} & Relevant\\ 
        & Repaired-Line           & {Class}    & Erroneous and Repaired & \\ \toprule 
        
        \multirow{2}{*}{1} & \code{d = (x-x1\rfill{) (}x-x1);} & $C_{32}$ \{$E_{15}$ \code{+*} \} 
            & \code{amount = \rfill{P (}1+T*R/100);} & \multirow{2}{*}{\cmark}\\
        & \code{d = (x-x1)\gfill{*}(x-x1);} & & \code{amount = P\gfill{*}(1+T*R/100);} \\\midrule

        \multirow{2}{*}{2} & \code{if (a\rfill{!(==)}0)} & $C_{148}$ \{$E_{1}$ \code{+!=\quad-!} \}  
            & \code{while (n \rfill{!} 0)} & \multirow{2}{*}{\cmark}\\
        & \code{if (a\gfill{!=}0)} & & \code{while (n \gfill{!=} 0)} & \\ \midrule
        
        \multirow{2}{*}{3} & \code{for(j=0\rfill{,} j<10\rfill{,} j++)} & $C_{10}$ \{$E_{7}$ \code{+;\quad-,} \}  
            & \code{for(i=0\rfill{,} i<n; i++)\{} & \multirow{2}{*}{\cmark}\\
        & \code{for(j=0\gfill{;} j<10\gfill{;} j++)} & & \code{for(i=0\gfill{;} i<n; i++)\{} & \\\midrule 

        \multirow{2}{*}{4} & \code{printf\rfill{[}"\%d", a\rfill{]};} & $C_{210}$ \{$E_{62}$ \code{+( +) -[ -]} \}  
            & \code{sort\rfill{[}rsum, n\rfill{]};} & \multirow{2}{*}{\cmark}\\
        & \code{printf\gfill{(}"\%d", a\gfill{)};} & & \code{sort\gfill{(}rsum, n\gfill{)};} & \\ \midrule             
        
        \multirow{2}{*}{5} & \code{printf("\%d", a)\rfill{)};} & $C_{40}$ \{$E_{19}$ \code{-)} \}  
            & \code{printf("(\%f,\%f"\rfill{)}, x,y);} & \multirow{2}{*}{\cmark}\\
        & \code{printf("\%d", a);} & & \code{printf("(\%f,\%f)", x,y);} & \\ \midrule

        \multirow{2}{*}{6} & \code{b=\rfill{xyz}+1;} & $C_{2}$ \{$E_{3}$ \code{+INT\quad-INVALID} \} 
            & \code{scanf("\%d", \&\rfill{a});} & \multirow{2}{*}{\cmark}\\        
        & \code{b=\gfill{a}+1;} & & \code{scanf("\%d", \&\gfill{n});} & \\\midrule
        

        \multirow{2}{*}{7} & \code{\rfill{p+'a' =} p;} & $C_{8}$ \{$E_{10}$ \code{+==\quad-=} \} 
            & \code{if (a \rfill{=} b)} & \multirow{2}{*}{\xmark}\\
        & \code{\gfill{p =} p+'a';} & & \code{if (a \gfill{==} b)} \\

        \bottomrule
        
    \end{tabular}     

\caption[Example generation on buggy programs]{Sample erroneous-repaired lines and the top erroneous-repaired code example suggested by \tegcer. The third column lists the top error-repair class predicted by \tegcer. The last column indicates whether the feedback is relevant to the student's error/repair.}
\label{table:eg_examples}    
\end{table*}    

\subsection{Unseen Errors}
The design of our error-repair class is flexible in accommodating new mistakes that students would make, unobserved in our training dataset. To further elaborate, consider an unlikely buggy line ``\code{printf("\%d" xyz!0);}'', where the student has made 3 different mistakes. Namely, a comma separator `\code{,}' is missing between both expressions, the inequality operator `\code{!=}' is incorrectly written as `\code{!}', and ``\code{xyz}'' is an undeclared variable of type ``\code{INVALID}''. The compiler reports just a single error $E_1$: \code{expected `)'} on this particular line. 

Our neural network classifier has never come across this combination of errors during its entire learning phase. Nonetheless, it is able to capture the errors better than a standard compiler, and predict relevant repair classes for all 3 mistakes individually. The top-4 repair classes predicted by \tegcer are $C_{148}$ \{\code{$E_1$ +!= -!} \}, $C_2$ \{\code{$E_3$ +INT -INVALID}\}, $C_{35}$ \{\code{$E_{12}$ +INT -INVALID}\}, and $C_7$ \{\code{$E_1$ +,} \}. Hence, \tegcer can be used to successfully generate relevant feedback on unseen errors by sampling multiple top-N classes. Notice that \tegcer does not restrict itself to the cryptic compiler reported error $E_1$, and this freedom allows it to match buggy program with relevant repair, by borrowing repairs observed in different errors $E_3$ and $E_{12}$.


\section{Example Suggestions}
\label{sec:eg_egSuggestion}


Given an erroneous program, \tegcer suggests relevant example erroneous-repaired pairs of another student. Either the erroneous code, or the repaired code, or both the code pairs can be provided to the student, depending on the instructor's preference. \tegcer's implementation is made publicly available~\footnote{\url{https://github.com/umairzahmed/tegcer}}. Table~\ref{table:eg_examples} lists few sample erroneous lines by students, and the top example feedback generated by \tegcer. 

In the first example listed in row $\#1$ of Table~\ref{table:eg_examples}, the student fails to realize that that an asterisk `\code{*}' operator is required between the two integer expressions. While in rows $\#2$ and $\#3$, the student has misunderstood the syntax of inequality operator `\code{!=}' and \code{for} loop statement specifier, respectively. The suggested examples by \tegcer reflects the same confusion and desired repair, as the original student's erroneous code. The compiler reports cryptic error messages \code{$E_{15}$: called object type "int" is not a function}, and \code{$E_{1}$: expected ')'}, for rows $\#1$ and $\#2$, respectively. Hence, providing \tegcer's examples as feedback to novice students could be more useful over these compiler messages, in understanding the error and its repair.

Rows $\#4$ and $\#5$ deal with bracketing issues in \code{printf} function calls, where a student has incorrectly used square brackets ``[ ]'' and added a spurious closing parenthesis ``\code{)}'', respectively. \tegcer's example of correct parenthesis usage in \code{printf} and a user-defined function call is highly relevant. 

Row $\#6$ lists an example of undeclared variable, where \tegcer suggests replacing it with a previously declared integer variable based on the erroneous line's local context.

Row $\#7$ demonstrates the limitation of our dense neural network, which relies on unigrams, bigrams and compilation error messages to predict the repair class. In this case, the erroneous line is \code{p+'a'=p;}, and the student is confused between the lvalue (left) and rvalue (right) of assignment operator. The input tokens passed to the neural network are \{\code{<ERR>}, \code{$E_{10}$}, \code{<UNI>}, \code{CHAR}, \code{+}, \code{'}, \code{LITERAL-CHAR}, \code{'}, \code{=}, \code{CHAR}, \code{;}, \code{<BI>}, \code{CHAR\_+}, \code{+\_'}, \code{'\_LITERAL-CHAR}, \code{LITERAL-CHAR\_'}, \code{'\_=}, \code{=\_CHAR}, \code{CHAR\_;}, \code{<EOS>}\}. 

As we can see, there is very little indication from these unigram and bigram tokens, that the issue is due to presence of multiple tokens in lvalue of assignment. In fact, examples belonging to a frequently occurring class, $C_8$, contain very similar unigram/bigram tokens, where the confusion is between usage of assignment ``\code{=}'' and equality ``\code{==}'' operator. Hence, the classifier incorrectly predicts the repair class $C_8$ \{\code{$E_{10}$ +== -=}\} for row $\#7$ source line, suggesting replacement of ``\code{=}'' with ``\code{==}''. This issue could be resolved by using more complex neural networks, which capture the entire sequence information. However, these would typically require larger training sample size for each individual class.

\section{User Study}

\begin{table*}[tbp] 
    \centering
    \begin{tabular}{llrr rr rr rr}
        \toprule
        \textbf{Semester} & \textbf{Feedback} & \textbf{\#Students} & \textbf{\#Human} & \textbf{\#Success} & \textbf{\#Failure} & \multicolumn{2}{c}{\textbf{Time-Taken (sec)}}  & \multicolumn{2}{c}{\textbf{\#Attempt}}\\ 
        & \textbf{Type} & & \textbf{Tutors} & \textbf{Compile} & \textbf{Compile} & \textbf{AVG} & \textbf{STD-DEV} & \textbf{AVG} & \textbf{STD-DEV} \\
        \midrule
        2017--2018--I  & Manual & 453 & 44 & 161,326 & 15,026 & 103  & 155 & 2.20 & 2.10   \\
        2017--2018--II & Example & 238 & 22 & 97,763 & 8,624 & 78  & 132 & 1.99 & 1.86\\
        \bottomrule
    \end{tabular}
    \caption{Success and failure of compilation-requests made by students during the 7-week labs across various offerings}
    \label{tab:eg_semester_students}
\end{table*}


In order to measure the pedagogical benefit of providing example based feedback, we deployed \tegcer during the 2017-2018-II fall semester offering of Introduction to C Programming (CS1) course at IIT Kanpur. 
This course was credited by more than 470 first year undergraduate students.
 A major component of the course was weekly programming assignments, termed \textit{labs}, which were attempted under the guidance of $40+$ post-graduate Teaching Assistants (TAs), over a period of $14$ weeks. Our analysis focuses on the first 7 weeks of the course, while \tegcer was deployed to guide these students instead of the TAs. Every week, the students were given 3 hours to complete assignments of varying difficulty. The theme of assignments covered during these 7 weeks was input-output, conditionals, iterations, nested-iterations, functions, arrays and matrices. All intermediate code attempts were recorded on Prutor~\cite{das2016prutor}, 
a custom IDE developed for teaching CS1.

During this 21 hour (3 hours x 7 weeks) period, \tegcer was deployed to provide live feedback to students, in addition to the error messages reported by compiler. Fig~\ref{fig:eg_ui} demonstrates the user-interface of \tegcer integrated with the Prutor IDE~\cite{das2016prutor}, 
and Fig~\ref{fig:eg_systemTrain} provides the backend system overview. Whenever a student in this course offering encountered a compilation error, we first ran our neural network classifier to predict the error-repair class that it belongs to. Then, we searched our student code repository of 2015-2016-II semester for code attempts that belong to the same error-repair class. Finally, the repaired line of these previous code attempts is extracted and presented to student as feedback, sequentially in decreasing order of the lines' frequency in our corpus. 

Initially, only the top frequent repaired example is shown. The student could request for additional examples using a \textit{More?} button, upto a maximum of $10$ per line, and master the programming syntax from multiple correct examples. For instance, in Fig~\ref{fig:eg_ui}, the student has requested two additional examples for line \#8, by utilizing the \textit{More?} option. 

Human tutors were asked to not help students resolve compilation errors, unless they were unable to resolve them with automated feedback for more than 5 minutes.

\subsection{Baseline Comparison}
The previous semester 2017-2018-I course offering of the same CS1 course at the same university (IIT Kanpur), attended by a different set of students, 
was used as a comparative baseline. Notably, the division of the first year student into groups that would attend CS1 in the Fall and Spring semester respectively was made administratively and randomly, using no student input. Thus, the two student populations could be treated as substantially identical. Also, both these course offerings followed the same syllabus and weekly lab settings. However, different sets of lab assignments were framed for both offerings. These students used the same browser based IDE to complete their programming tasks, but without access to any feedback tool. They relied on the compiler error messages and manual help by human tutors to resolve their compilation errors.

\subsection{Overall Results}

We compared student performance on two different metrics, namely \textit{\#attempts} and \textit{time-taken}. 
\textit{Time-Taken} is defined as the time elapsed from the first occurrence of a compilation error, to the next successful compilation attempt made by student.  \textit{\#Attempts} is defined as the number of unsuccessful compilation requests made by the student before finally resolving the error. For both the TEGCER and the baseline group, we filtered out instances where students took more than 15 minutes time to resolve their compilation error, to discount intrusions such as being away from the desk or doing something else.

Table~\ref{tab:eg_semester_students} shows the results of $453$ students from the first 7 weeks of the baseline semester and $238$ students from the first 7 weeks of the semester with access to \tegcer. Students in our control group took $103$ seconds on an average to resolve any compilation error, with manual help provided by TAs when they get stuck. Significantly lower $78$ seconds on average were required by our experimental group, even accounting for delayed manual help by TAs on getting stuck on a compilation error. A Welch's t-test for testing statistical significance of variance in average \textit{time-taken} (log transformed), returned $t=16.63, p < 0.0001$. The test is robust for  unequal sample sizes as it is in our case. Similarly, the average \#attempts of $1.99$ required by experimental group is significantly lower compared to the $2.20$ required by the baseline group (Welch's t-test $t=8.06, p < 0.0001$). 

\subsection{\tegcer helps more when problems are harder}

\begin{figure}[tbp] 
    \centering        
    \includegraphics[width=0.95\linewidth, keepaspectratio]{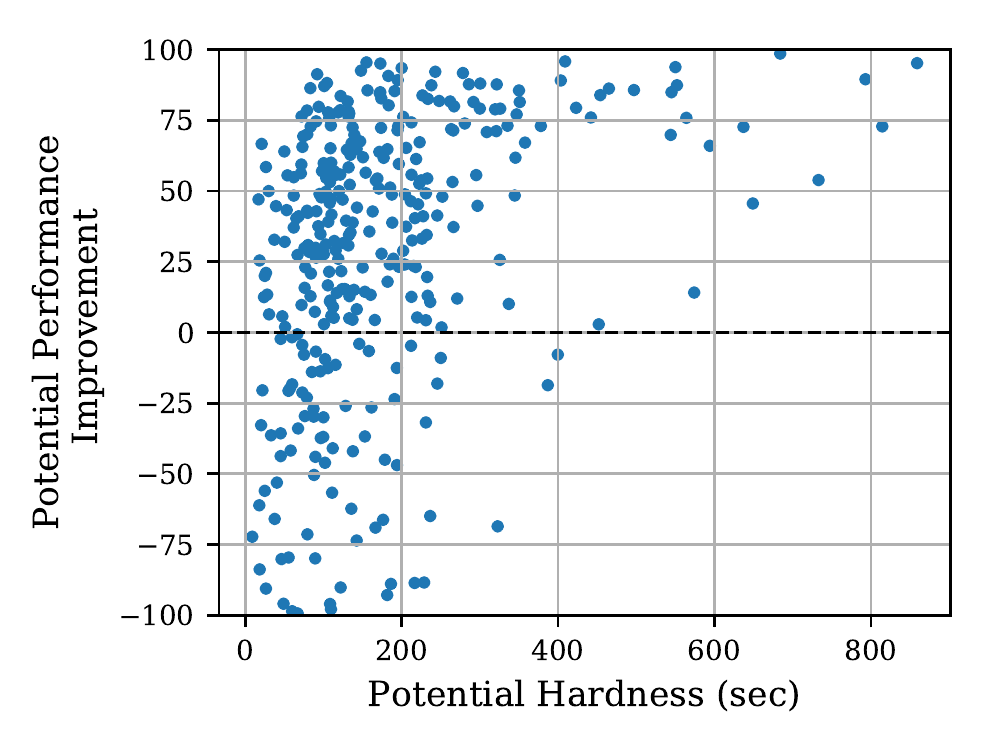}
    \caption{Potential hardness (x-axis) of resolving an error-group plotted against potential performance improvement (y-axis) offered by our example generation tool.}
    \label{fig:eg_userStudy}
\end{figure}

Due to the heavily imbalanced distribution in  compilation error types, the most frequent ones tend to dominate the overall average result. Hence, we also compare time-taken across individual error groups ($EG$). A total of $405$ unique $EG$s were encountered common to both semesters. In $253$ ($62.5\%$) of these error-groups, the experimental group with access to \tegcer resolved their errors faster, compared to $152$ ($37.5\%$) cases where the baseline group has a lower average time.

Figure~\ref{fig:eg_userStudy} plots the potential hardness of each $EG$, measured in terms of average time-taken by baseline students ($\text{time}_{\text{b}}$) to resolve programs belonging to a particular error type ($EG$), against the potential performance improvement offered by \tegcer, measured as the normalized difference between the time-taken by both cohorts $(\text{time}_{\text{b}} - \text{time}_{\text{e}})/\text{time}_{\text{b}}$, for the same error type. Note that, each point in the figure represents one $EG$ encountered in both semesters. From this figure it is evident that, while errors that are easier to resolve may or may not benefit from \tegcer, it is predominantly more helpful than human TAs for more difficult errors. This is a crucial observation, as it brings out the value of deploying a comprehensive tool like \tegcer as opposed to using simpler solutions, such as explaining compiler error messages better.

\subsection{Interesting Error Groups}

\begin{figure}[t]
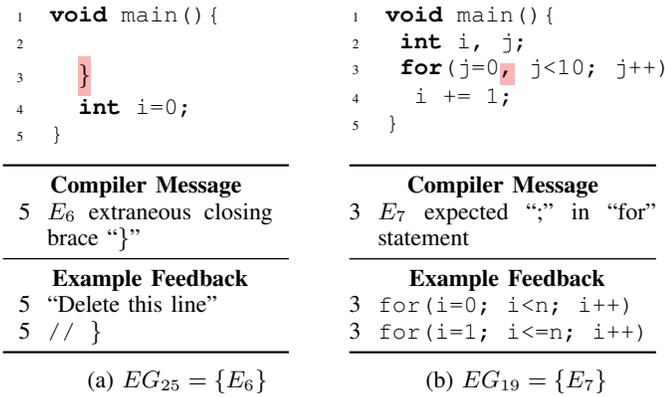
 
\centering
\mbox{ 
    \begin{subfigure}[t]{.49\linewidth}
        \centering \small
\begin{lstlisting}[]
void main(){
    
  $\text{\colorbox{red!30}{\!\!\}\!\!}}$
  int i=0;		
}
\end{lstlisting}
    	
    	\hspace{-3em}
    	\begin{tabular}{l@{\ \ }p{3cm}}
            \toprule
            \multicolumn{2}{c}{\textbf{Compiler Message}}\\ 
            5 & $E_{6}$  {extraneous closing brace ``\}''} \\ 
            \midrule
            
            \multicolumn{2}{c}{\textbf{ Example Feedback}}\\
            5 & {``Delete this line''} \\
            5 & \texttt{// \}} \\
            \bottomrule
        \end{tabular}
    
    	\caption{$EG_{25} = \{E_6\}$}
    	\label{fig:error7}
    \end{subfigure}	%
    
    \begin{subfigure}[t]{.5\linewidth}
        \centering \small
\begin{lstlisting}[]
void main(){
	int i, j;
	for(j=0$\text{\colorbox{red!30}{\!\!,\!\!}}$ j<10; j++)
		i += 1;
}
\end{lstlisting}
    	\vspace{.45em}
    	\hspace{-1.5em}
    	\begin{tabular}{@{}l@{\ \ }p{3.7cm}@{}}
            \toprule
            \multicolumn{2}{c}{\textbf{Compiler Message}}\\ 
            3 & $E_{7}$ {expected ``;'' in ``for'' statement} \\
            \midrule
            
            \multicolumn{2}{c}{\textbf{ Example Feedback}}\\
            3 & \texttt{for(i=0; i<n; i++)} \\
            3 & \texttt{for(i=1; i<=n; i++)} \\
            \bottomrule
        \end{tabular}
    	\caption{$EG_{19} = \{E_7\}$}
    	\label{fig:error19}
    \end{subfigure}	
}
\caption{Sample programs and top-2 example suggestions, for (a) $EG_{25}$ where students with access to \tegcer perform poorly, (b) $EG_{19}$ where examples seems to work better.}
\label{fig:sample_codes_feedback}
\end{figure}

From our earlier analysis, it is apparent that automated example based feedback generated by \tegcer does help students resolve compilation errors faster on average. In this section, we highlight two particular cases wherein \tegcer's feedback helped the most (or least) compared to our baseline.


Students with access to \tegcer feedback seem to perform worse than baseline group of students when \tegcer produces an irrelevant feedback. Consider the error group $EG_{25}$ consisting of $E_{6}$, which is typically encountered when students inadvertently add extra closing brace ``\}''. Fig~\ref{fig:error7} shows a simple code which triggers this errors, along with the compiler message and \tegcer's feedback. Since \tegcer relies on compiler for error localization, it focuses its feedback on line \#5. Hence, \tegcer incorrectly suggests deleting the brace on line \#5 to  successfully resolve the compilation error, instead of suggesting the more relevant fix of deleting spurious ``\}'' on line \#3. This seems to adversely affect students with access to \tegcer who take 84 seconds on average to resolve $E_6$ errors, compared to our baseline group of students who require 46 seconds on average.


In contrast, for $EG_{19}$, \tegcer assistance shows much faster error resolution (46 seconds on average) than the baseline students (80 seconds on average). A simple example code belonging to this error group is shown in Fig~\ref{fig:error19}, where the student is confused with for-loop syntax. While the compiler message correctly suggests the replacement of comma ``,'' with a semi-colon ``;'' after loop initialization, \tegcer suggests the same through examples. Its top-2 examples demonstrate the 0-to-(N-1) and 1-to-N loop iteration. On requesting further examples, \tegcer's examples suggests different ways of writing for loops with empty-initialization, empty-condition or empty-increment mode, which is perhaps helping students master the for-loop syntax better.

    \section{Related Work}
Compiler error messages are the quickest form of feedback provided to a novice programmer. A novice programmer can be expected to learn better if error messages are augmented with the information to fix them. Improving the compiler error messages and repairing them has been active areas of research over last several decades. We highlight some recent related work targeting introductory programmers in this section.


\subsection{Compilation Error Repair}
sk\_p~\cite{pu2016sk_p} uses sequence-to-sequence (Seq2Seq) network to generate a  statement repair, based on other statements observed around it (local context). DeepFix~\cite{gupta2017deepfix} trains a Recurrent Neural Network (RNN) on artificially generated correct programs to jointly perform error localization and correction on entire buggy program. \tracer~\cite{ahmed2018compilation} improves upon DeepFix by abstracting variables with their types, and training a Seq2Seq on erroneous-repaired pairs of abstract program lines. RLAssist~\cite{gupta2018deep} treats the entire buggy program as a 2D game, and trains a reinforcement-learning algorithm which utilizes movement and edit based actions to improve upon DeepFix.
These compilation error repair tools typically use some form of generative deep networks. They achieve low repair accuracy in the range of 20-45\% since the tools have to predict not only the set of repair tokens to add/remove, but also their exact arrangement for the program to compile. In contrast, \tegcer achieves high \acci{1} accuracy of $87\%$ by utilizing a dense classifier of error-repair tokens instead of generative model.

\subsection{Compilation Error Message Enhancement}
Denny \etal~\cite{denny2014enhancing} and Parihar \etal~\cite{gradeit} enhance the compiler generated error messages to make it easy to comprehend for novices. Whenever a submission triggers compilation error, these system show the enhanced error message corresponding to the error, possibly accompanied by an explanation and an example of correct code. However, in these work the enhancements and the example are created statically, unlike \tegcer\ where the examples are mined from old submissions.

Becker \etal~\cite{becker2018effects} have recently showed that enhanced error messages are effective in error repair by novice programmers. In contrast, other studies have found that additional information does not help students significantly~\cite{denny2014enhancing, pettit2017enhanced}, and that placement and structuring of the message is more crucial~\cite{nienaltowski2008compiler}. This confusion in the literature suggests that further research is clearly needed to establish the optimal methods for presenting error messages to students. Our large sample study's observations should be helpful in this regard.

Barik \etal~have developed a large number of guidelines about how the error messages should be presented to developers~\cite{barik2014compiler,barik2017do,barik2018should,barik2018error}. Most of these guidelines could be applicable to novice programmers as well, for example, the recommendation that the error message which appears in the editor must be close to the source of error.

\subsection{Compilation Error Feedback Generation}

The closest related work to ours is HelpMeOut~\cite{hartmann2010helpmeout}, which relies on manual contribution by students to record code examples and explanations. HelpMeOut ranks the relevant examples based on the error similarity and user provided ratings, while \tegcer ranks the examples based on the similarity of error and predicted repair desired by student. 

Other related work that use machine learning include the work by Wu \etal~\cite{wu2017learning} and Santos \etal~\cite{santos2018syntax}. The technique by Wu \etal~\cite{wu2017learning} is similar to ours wherein they apply machine learning techniques on a large corpus of erroneous student programs, albeit focusing on type related errors. This learnt model is used to generate high quality error message and repair suggestions at the expression level. Santos \etal~\cite{santos2018syntax} differ from \tegcer\ as they use   correct programs by experienced programmers from GitHub\footnote{https://github.com} instead of prior student submissions.

Thiselton \etal~propose PYCEE~\cite{thiselton2019enhancing}, an IDE plugin that queries Stack Overflow\footnote{https://stackoverflow.com} automatically whenever a programmer encounters a Python error message. The Stack Overflow answers are summarized and presented as enhanced error message within the IDE.
They report on the usefulness of 115 compiler error message enchancements, encountered during a user study of 16 programmers using PYCEE.

    \section{Threats to Validity}
In this section, we describe some threats to the validity of our study. \textit{External validity} threats relate to the generalizability of the results. In this study we had over 600 students from two semesters for a period of 7 weeks. We have conducted the study on a sample size that represents an educational setting with novice programmers. \textit{Internal validity} threat relates to the environment under which experiments were carried out. 

Our baseline participants where students had access to teaching assistants only were selected from a different semester (2017--2018--I), compared to our control group participants with access to \tegcer (2017--2018--II). While it is desirable to randomize students from the same semester, our choice of study design was ethically constrained by the prospect of one group receiving unfair advantage over the other in the live graded course offering.

During the measurement interval, issues such as mortality (students withdrawing from the lab during data collection) and maturation (students changing their characteristics by parameters outside of the study) did not arise. \textit{Construct validity} relates to measurement of variables used during the study. We have used standard dependent variables such as time-taken to resolve compiler error(s), \textit{\#attempts}. Our metric of time-taken to resolve errors does not take into account time spent away from the task, or spent resolving logical errors. To partially account for time away from task problems, we discounted instances of errors persisting for longer than 15 minutes. Since we are unable to remove all possible sources of variation from our variable, it is appropriate to consider the possibility that our quantitative predictions may have less than ideal veridicality while remaining directionally correct.

\section{Conclusion}
\label{sec:eg_concl}
We presented \tegcer, an automated example generation tool. It is based on the realization that while students make multiple types of compilation errors on various programming assignments, the kind of repairs performed by them is limited to addition/deletion of few abstract program tokens. To this end, we first collated and labeled a large dataset of $15,000$+ aligned erroneous-repaired student program pairs, with $212$ unique error-repair classes ($C$). These classes capture the combination of errors ($E$s) made by students, and the repairs ($R$s) attempted by them. We aim to release this labelled dataset into the public domain to aid further research.

This novel labelling, data pre-processing steps of error localization and abstraction, enabled a simple dense neural network with a single hidden layer achieve \acci{3} of $97.68\%$ on a challenging task of predicting a correct label from $212$ different ones. 
We conducted a large-scale empirical evaluation comparing error repair times for a cohort of about 250 students programming with the aid of \tegcer in a controlled lab environment with those for a similar cohort of about 450 students programming in the same environment with human assistance. We found that example based feedback from \tegcer helped resolve errors 25\% faster on average. Further, \tegcer was consistently more useful for fixing harder errors. The large scale and controlled nature of our empirical evaluation allows us to assert with high confidence that assistance by \tegcer is comparable to, if not better than, the assistance provided by human teaching assistants in lab settings.

    \section*{Acknowledgements}
    U. Z. Ahmed was supported by an IBM PhD fellowship for the period 2017--2018.


\clearpage
\bibliographystyle{IEEEtran}
\balance
\bibliography{main}

\begin{thebibliography}{10}
\providecommand{\url}[1]{#1}
\csname url@samestyle\endcsname
\providecommand{\newblock}{\relax}
\providecommand{\bibinfo}[2]{#2}
\providecommand{\BIBentrySTDinterwordspacing}{\spaceskip=0pt\relax}
\providecommand{\BIBentryALTinterwordstretchfactor}{4}
\providecommand{\BIBentryALTinterwordspacing}{\spaceskip=\fontdimen2\font plus
\BIBentryALTinterwordstretchfactor\fontdimen3\font minus
  \fontdimen4\font\relax}
\providecommand{\BIBforeignlanguage}[2]{{%
\expandafter\ifx\csname l@#1\endcsname\relax
\typeout{** WARNING: IEEEtran.bst: No hyphenation pattern has been}%
\typeout{** loaded for the language `#1'. Using the pattern for}%
\typeout{** the default language instead.}%
\else
\language=\csname l@#1\endcsname
\fi
#2}}
\providecommand{\BIBdecl}{\relax}
\BIBdecl

\bibitem{ucbClass}
A.~Mabanta, C.~Hunt, and S.~Najmadadi, ``How big is uc berkeley’s biggest
  class?''
  \url{https://web.archive.org/web/20171226115213/http://www.dailycal.org/2013/09/03/how-big-is-uc-berkeleys-biggest-class/},
  accessed: 2017-12-26.

\bibitem{udacityClass}
K.~Mangan, ``A first for udacity,''
  \url{https://web.archive.org/web/20171202001557/http://www.chronicle.com/article/A-First-for-Udacity-Transfer/134162/},
  accessed: 2017-12-02.

\bibitem{camp2015booming}
T.~Camp, S.~Zweben, E.~Walker, and L.~Barker, ``Booming enrollments: Good
  times?'' in \emph{Proceedings of the 46th ACM Technical Symposium on Computer
  Science Education}, ser. SIGCSE '15.\hskip 1em plus 0.5em minus 0.4em\relax
  ACM, 2015, pp. 80--81.

\bibitem{mccauley2008debugging}
R.~McCauley, S.~Fitzgerald, G.~Lewandowski, L.~Murphy, B.~Simon, L.~Thomas, and
  C.~Zander, ``Debugging: a review of the literature from an educational
  perspective,'' \emph{Computer Science Education}, vol.~18, no.~2, 2008.

\bibitem{becker2019unexpected}
B.~A. Becker, P.~Denny, R.~Pettit, D.~Bouchard, D.~J. Bouvier, B.~Harrington,
  A.~Kamil, A.~Karkare, C.~McDonald, P.-M. Osera, J.~L. Pearce, and J.~Prather,
  ``Unexpected tokens: A review of programming error messages and design
  guidelines for the future,'' in \emph{Proceedings of the 2019 ACM Conference
  on Innovation and Technology in Computer Science Education}, ser. ITiCSE
  '19.\hskip 1em plus 0.5em minus 0.4em\relax ACM, 2019, pp. 253--254.

\bibitem{becker2018effects}
B.~A. Becker, K.~Goslin, and G.~Glanville, ``The effects of enhanced compiler
  error messages on a syntax error debugging test,'' in \emph{Proceedings of
  the 49th ACM Technical Symposium on Computer Science Education}, ser. SIGCSE
  '18.\hskip 1em plus 0.5em minus 0.4em\relax ACM, 2018, pp. 640--645.

\bibitem{monperrus2015automatic}
M.~Monperrus, ``{Automatic Software Repair: a Bibliography},'' \emph{University
  of Lille, Tech. Rep. hal-01206501}, 2015.

\bibitem{ahmed2018compilation}
U.~Z. Ahmed, P.~Kumar, A.~Karkare, P.~Kar, and S.~Gulwani, ``Compilation error
  repair: For the student programs, from the student programs,'' in
  \emph{Proceedings of the 40th International Conference on Software
  Engineering: Software Engineering Education and Training}, ser. ICSE-SEET
  '18.\hskip 1em plus 0.5em minus 0.4em\relax ACM, 2018, pp. 78--87.

\bibitem{lahtinen2005}
E.~Lahtinen, K.~Ala-Mutka, and H.-M. J\"{a}rvinen, ``A study of the
  difficulties of novice programmers,'' in \emph{Proceedings of the 10th Annual
  SIGCSE Conference on Innovation and Technology in Computer Science
  Education}, ser. ITiCSE '05.\hskip 1em plus 0.5em minus 0.4em\relax ACM,
  2005, pp. 14--18.

\bibitem{sweller1999}
J.~Sweller, \emph{\BIBforeignlanguage{English}{Instructional design}}.\hskip
  1em plus 0.5em minus 0.4em\relax ACER Press Camberwell, Vic, 1999.

\bibitem{gupta2018deep}
R.~Gupta, A.~Kanade, and S.~K. Shevade, ``Deep reinforcement learning for
  syntactic error repair in student programs,'' in \emph{The Thirty-Third
  {AAAI} Conference on Artificial Intelligence, {AAAI} 2019}, 2019, pp.
  930--937.

\bibitem{gupta2017deepfix}
R.~Gupta, S.~Pal, A.~Kanade, and S.~Shevade, ``Deepfix: Fixing common c
  language errors by deep learning,'' in \emph{Proceedings of the Thirty-First
  AAAI Conference on Artificial Intelligence}, ser. AAAI '17.\hskip 1em plus
  0.5em minus 0.4em\relax AAAI Press, 2017, pp. 1345--1351.

\bibitem{bhatia2016automated}
S.~Bhatia and R.~Singh, ``{Automated Correction for Syntax Errors in
  Programming Assignments using Recurrent Neural Networks},'' in \emph{2nd
  Indian Workshop on Machine Learning}, ser. IWML '16, 2016.

\bibitem{nienaltowski2008compiler}
M.-H. Nienaltowski, M.~Pedroni, and B.~Meyer, ``Compiler error messages: What
  can help novices?'' in \emph{Proceedings of the 39th SIGCSE Technical
  Symposium on Computer Science Education}, ser. SIGCSE '08.\hskip 1em plus
  0.5em minus 0.4em\relax ACM, 2008, pp. 168--172.

\bibitem{das2016prutor}
\BIBentryALTinterwordspacing
R.~Das, U.~Z. Ahmed, A.~Karkare, and S.~Gulwani, ``{Prutor: A System for
  Tutoring CS1 and Collecting Student Programs for Analysis},'' 2016,
  arXiv:1608.03828 [cs.CY]. [Online]. Available:
  \url{https://www.cse.iitk.ac.in/users/karkare/prutor/}
\BIBentrySTDinterwordspacing

\bibitem{lattner2004llvm}
C.~Lattner and V.~Adve, ``{LLVM}: A compilation framework for lifelong program
  analysis \& transformation,'' in \emph{Proceedings of the International
  Symposium on Code Generation and Optimization: Feedback-directed and Runtime
  Optimization}, ser. CGO '04.\hskip 1em plus 0.5em minus 0.4em\relax IEEE
  Computer Society, 2004, pp. 75--.

\bibitem{renkl2014toward}
A.~Renkl, ``Toward an instructionally oriented theory of example-based
  learning,'' \emph{Cognitive science}, vol.~38, no.~1, 2014.

\bibitem{hartmann2010helpmeout}
B.~Hartmann, D.~MacDougall, J.~Brandt, and S.~R. Klemmer, ``What would other
  programmers do: Suggesting solutions to error messages,'' in
  \emph{Proceedings of the SIGCHI Conference on Human Factors in Computing
  Systems}, ser. CHI '10.\hskip 1em plus 0.5em minus 0.4em\relax ACM, 2010, pp.
  1019--1028.

\bibitem{clara}
S.~Gulwani, I.~Radi\v{c}ek, and F.~Zuleger, ``Automated clustering and program
  repair for introductory programming assignments,'' in \emph{Proceedings of
  the 39th ACM SIGPLAN Conference on Programming Language Design and
  Implementation}, ser. PLDI '18.\hskip 1em plus 0.5em minus 0.4em\relax ACM,
  2018, pp. 465--480.

\bibitem{denny2012all}
P.~Denny, A.~Luxton-Reilly, and E.~Tempero, ``All syntax errors are not
  equal,'' in \emph{Proceedings of the 17th ACM Annual Conference on Innovation
  and Technology in Computer Science Education}, ser. ITiCSE '12.\hskip 1em
  plus 0.5em minus 0.4em\relax ACM, 2012, pp. 75--80.

\bibitem{chollet2015keras}
F.~Chollet \emph{et~al.}, ``Keras,'' \url{https://keras.io}, 2015.

\bibitem{rajaraman_ullman_2011}
A.~Rajaraman and J.~D. Ullman, \emph{Data Mining}.\hskip 1em plus 0.5em minus
  0.4em\relax Cambridge University Press, 2011.

\bibitem{allamanis2018survey}
M.~Allamanis, E.~T. Barr, P.~Devanbu, and C.~Sutton, ``A survey of machine
  learning for big code and naturalness,'' \emph{ACM Comput. Surv.}, vol.~51,
  no.~4, pp. 81:1--81:37, Jul. 2018.

\bibitem{hochreiter1997long}
S.~Hochreiter and J.~Schmidhuber, ``Long short-term memory,'' \emph{Neural
  Comput.}, vol.~9, no.~8, pp. 1735--1780, Nov. 1997.

\bibitem{krizhevsky2012imagenetCNN}
A.~Krizhevsky, I.~Sutskever, and G.~E. Hinton, ``Imagenet classification with
  deep convolutional neural networks,'' in \emph{Proceedings of the 25th
  International Conference on Neural Information Processing Systems - Volume
  1}, ser. NIPS '12.\hskip 1em plus 0.5em minus 0.4em\relax Curran Associates
  Inc., 2012, pp. 1097--1105.

\bibitem{srivastava2014dropout}
N.~Srivastava, G.~Hinton, A.~Krizhevsky, I.~Sutskever, and R.~Salakhutdinov,
  ``Dropout: A simple way to prevent neural networks from overfitting,''
  \emph{J. Mach. Learn. Res.}, vol.~15, no.~1, pp. 1929--1958, Jan. 2014.

\bibitem{abadi2016tensorflow}
M.~Abadi, P.~Barham, J.~Chen, Z.~Chen, A.~Davis, J.~Dean, M.~Devin,
  S.~Ghemawat, G.~Irving, M.~Isard, M.~Kudlur, J.~Levenberg, R.~Monga,
  S.~Moore, D.~G. Murray, B.~Steiner, P.~Tucker, V.~Vasudevan, P.~Warden,
  M.~Wicke, Y.~Yu, and X.~Zheng, ``Tensorflow: A system for large-scale machine
  learning,'' in \emph{Proceedings of the 12th USENIX Conference on Operating
  Systems Design and Implementation}, ser. OSDI '16.\hskip 1em plus 0.5em minus
  0.4em\relax USENIX Association, 2016, pp. 265--283.

\bibitem{kingma2014adam}
D.~P. Kingma and J.~Ba, ``Adam: {A} method for stochastic optimization,'' in
  \emph{3rd International Conference on Learning Representations}, ser. ICLR
  '15, 2015.

\bibitem{pu2016sk_p}
Y.~Pu, K.~Narasimhan, A.~Solar-Lezama, and R.~Barzilay, ``sk\_p: a neural
  program corrector for moocs,'' in \emph{Companion Proceedings of the 2016 ACM
  SIGPLAN International Conference on Systems, Programming, Languages and
  Applications: Software for Humanity}, 2016.

\bibitem{denny2014enhancing}
P.~Denny, A.~Luxton-Reilly, and D.~Carpenter, ``Enhancing syntax error messages
  appears ineffectual,'' in \emph{Proceedings of the 19th Conference on
  Innovation and Technology in Computer Science Education}, ser. ITiCSE
  '14.\hskip 1em plus 0.5em minus 0.4em\relax ACM, 2014, pp. 273--278.

\bibitem{gradeit}
S.~Parihar, Z.~Dadachanji, P.~K. Singh, R.~Das, A.~Karkare, and
  A.~Bhattacharya, ``Automatic grading and feedback using program repair for
  introductory programming courses,'' in \emph{Proceedings of the 22nd ACM
  Conference on Innovation and Technology in Computer Science Education}, ser.
  ITiCSE '17.\hskip 1em plus 0.5em minus 0.4em\relax ACM, 2017, pp. 92--97.

\bibitem{pettit2017enhanced}
R.~S. Pettit, J.~Homer, and R.~Gee, ``Do enhanced compiler error messages help
  students?: Results inconclusive.'' in \emph{Proceedings of the 2017 ACM
  SIGCSE Technical Symposium on Computer Science Education}.\hskip 1em plus
  0.5em minus 0.4em\relax ACM, 2017, pp. 465--470.

\bibitem{barik2014compiler}
T.~Barik, J.~Witschey, B.~Johnson, and E.~Murphy-Hill, ``{Compiler Error
  Notifications Revisited: An Interaction-First Approach for Helping Developers
  More Effectively Comprehend and Resolve Error Notification},'' in
  \emph{Companion Proceedings of the Conference on Software Engineering, ICSE},
  2014, pp. 536--539.

\bibitem{barik2017do}
T.~Barik, J.~Smith, K.~Lubick, E.~Holmes, J.~Feng, E.~Murphy-Hill, and
  C.~Parnin, ``Do developers read compiler error messages?'' in
  \emph{Proceedings of the 39th International Conference on Software
  Engineering}, ser. ICSE '17.\hskip 1em plus 0.5em minus 0.4em\relax IEEE
  Press, 2017, pp. 575--585.

\bibitem{barik2018should}
T.~Barik, D.~Ford, E.~Murphy-Hill, and C.~Parnin, ``How should compilers
  explain problems to developers?'' in \emph{Proceedings of the 2018 26th ACM
  Joint Meeting on European Software Engineering Conference and Symposium on
  the Foundations of Software Engineering}, ser. ESEC/FSE '18.\hskip 1em plus
  0.5em minus 0.4em\relax ACM, 2018, pp. 633--643.

\bibitem{barik2018error}
\BIBentryALTinterwordspacing
T.~Barik, ``{Error Messages as Rational Reconstructions},'' Ph.D. dissertation,
  North Carolina State University, 2018. [Online]. Available:
  \url{https://repository.lib.ncsu.edu/handle/1840.20/35439}
\BIBentrySTDinterwordspacing

\bibitem{wu2017learning}
B.~Wu, J.~P. Campora~III, and S.~Chen, ``Learning user friendly type-error
  messages,'' \emph{Proc. ACM Program. Lang.}, vol.~1, no. OOPSLA, pp.
  106:1--106:29, Oct. 2017.

\bibitem{santos2018syntax}
E.~A. Santos, J.~C. Campbell, D.~Patel, A.~Hindle, and J.~N. Amaral, ``Syntax
  and sensibility: Using language models to detect and correct syntax errors,''
  in \emph{2018 IEEE 25th International Conference on Software Analysis,
  Evolution and Reengineering (SANER)}.\hskip 1em plus 0.5em minus 0.4em\relax
  IEEE, 2018, pp. 311--322.

\bibitem{thiselton2019enhancing}
E.~Thiselton and C.~Treude, ``{Enhancing Python Compiler Error Messages via
  Stack Overflow},'' in \emph{Proceedings of the 19th International Symposium
  on Empirical Software Engineering and Measurement}, ser. ESEM '19, Jun 2019,
  p. arXiv:1906.11456.

\end{thebibliography}

\end{document}